%% file: main.tex
\PassOptionsToPackage{noend}{algorithmic}
\PassOptionsToPackage{dvipsnames}{xcolor}
\PassOptionsToPackage{breakspaces}{clevethm}
\PassOptionsToPackage{hypertexnames=false}{hyperref}
\PassOptionsToPackage{capitalize,nameinlink}{cleveref}
\PassOptionsToPackage{textsize=footnotesize}{todonotes}
\documentclass[journal,10pt,twoside]{IEEEtranTCOM}
\usepackage{graphics}
\usepackage{adjustbox}
\usepackage[outdir=./images/]{epstopdf}
\usepackage{caption}
\usepackage{subcaption}
\usepackage{xspace}
\usepackage{cite}
\usepackage{float}
\usepackage{pgf}
\usepackage{tikz}
\usepackage{amsfonts}
\let\olddddot\dddot
\let\oldddddot\ddddot
\usepackage{accents}
\let\dddot\olddddot
\let\ddddot\oldddddot

\normalsize
\input{TeX/Preamble.tex}

\title{Cooperative Uplink Channel Estimation in User-Centric Cell-free Massive MIMO Communication Networks}
\author{Pourya Behmandpoor, 
Marc Moonen
\thanks{\ThanksToMarc\\\Affiliation}
}

\begin{document}
\maketitle

    \input{TeX/Text/abstract.tex}

    \section{Introduction}\label{sec:intro}
        \input{TeX/Text/introduction.tex}
        
    \section{System Model}
        \subsection{Network Model}
            \input{TeX/Text/nework_model.tex}
        \subsection{Scalable AP Assignment}\label{sec:AP_assignment}
            \input{TeX/Text/user_centeric_model.tex}
        \subsection{Channel Model}\label{sec:channel_model}
            \input{TeX/Text/channel_model.tex}

    \section{Distributed Adaptive Node-specific Signal Estimation}\label{sec:DANSE}
        \input{TeX/Text/DANSE.tex}

    \section{Proposed Cooperative Channel Estimation}\label{sec:CE}
        \input{TeX/Text/channel_estimation.tex}

    \section{Numerical Experiments} \label{sec:sim}
        \input{TeX/Text/simulations.tex}

    \section{Conclusion}
        \input{TeX/Text/conclusion.tex}

\appendices
\bibliographystyle{IEEEtranTCOM}
\bibliography{IEEEabrv,TeX/CFmMIMO.bib}

\end{document}

%% file: TeX/Preamble.tex
\usepackage{booktabs}
\usepackage{makecell}
\usepackage{float}
\usepackage{algorithmic}
\usepackage{bbm}
\usepackage{letltxmacro}
\usepackage[most]{tcolorbox}
\usepackage[%
	thmreset=section,
	eqreset=section,
	noload={algpseudocode,algorithmic}
]{myPreamble}

\pgfplotsset{compat=1.16}
\allowdisplaybreaks

\newlength\myindent
\usepackage[T3,T1]{fontenc}
\DeclareSymbolFont{tipa}{T3}{cmr}{m}{n}
\DeclareMathAccent{\invbreve}{\mathalpha}{tipa}{16}

\newcommand{\mathcirc}{\accentset{\circ}}

\newcommand{\C}{\mathbb{C}}

\DeclareMathOperator{\blkdiag}{blockdiag}

\makeatletter
	\newcommand{\E}{\@ifstar\@@E\@E}					
	\newcommandx{\@E}[3][1={},3={}]{
	\mathbb E_{#1}\ifstrempty{#2}{}{
		\left[
			#2\ifstrempty{#3}{}{\mid #3}
		\right]
	}
	}
	\newcommandx{\@@E}[3][1=k,3={}]{
	\mathbb E_{#1}\ifstrempty{#2}{}{
		[#2\ifstrempty{#3}{}{|#3}]
	}
	}
\makeatother

\Crefname{ALC@unique}{Line}{Lines}
\renewcommand{\algfont}{\bf}

\newlist{algsubstates}{enumerate}{2}
\makeatletter
	\setlist[algsubstates,1]{
		label={\alph*:},
		ref={\theALC@line.\alph*},
		itemsep=0pt,
		partopsep=0pt,
		topsep=0pt,
		parsep=0pt,
        leftmargin=12pt
	}
\makeatother
\newcommand{\fixstate}{\addtocounter{ALC@line}{-1}\refstepcounter{ALC@line}\phantomsection}
\newcommand{\State}{\STATE\fixstate}

\makeatletter
	\renewcommand\theALC@line{{\oldstylenums\arabic{ALC@line}}}
	\def\theHALC@line{\thealgorithm-\arabic{ALC@line}}
\makeatother
\crefname{ALC@line}{step}{steps}
\Crefname{ALC@line}{Step}{Steps}
\crefalias{algsubstatesi}{ALC@line}

\newlist{tblsubstates}{enumerate}{2}
\makeatletter
	\setlist[tblsubstates,1]{
		label={\alph*:},
		ref={\theALC@line.\alph*},
		itemsep=0pt,
		partopsep=0pt,
		topsep=0pt,
		parsep=0pt,
	}
\makeatother
\crefname{TBL@line}{table}{tabless}
\Crefname{TBL@line}{table}{Tabless}
\crefalias{tblsubstatesi}{TBL@line}

\newenvironment{talign*}
 {\let\displaystyle\textstyle\csname align*\endcsname}
 {\endalign}

 \newcommand{\Affiliation}{%
 Pourya Behmandpoor is with Vrije Universiteit Brussel (VUB), Department of Electronics and Informatics (ETRO), Pleinlaan 2, B-1050 Brussels, Belgium (e-mail: pourya.behmandpoor@vub.be). Marc Moonen is with KU Leuven, Department of Electrical Engineering (ESAT), STADIUS Center for Dynamical Systems, Signal Processing and Data Analytics (e-mail: marc.moonen@esat.kuleuven.be).
 }

 \newcommand{\ThanksToMarc}{This research work was carried out at the ESAT Laboratory of KU Leuven, in the frame of Research Project FWO nr. G0C0623N 'User-centric distributed signal processing algorithms for next generation cell-free massive MIMO based wireless communication networks'. The scientific responsibility is assumed by its authors.}

%% file: TeX/Text/abstract.tex
\begin{abstract}
    Cell-free massive multi-input-multi-output (CFmMIMO) communication networks aim to provide uniform quality of service by distributing access points (APs) across a coverage area. In \emph{user-centric} variants, each user equipment (UE) can choose a \emph{cluster} of APs with the best channel conditions (\eg the closest APs) for accessing service. This approach eliminates the notion of cells with dedicated regions and APs, as found in cellular mMIMO communication networks. 
    Estimating uplink channels between UEs and APs is a crucial step in CFmMIMO communication networks; however, existing channel estimation (CE) approaches typically originate from mMIMO systems without considering the unique properties of CFmMIMO communication networks.  
    For instance, shorter AP-UE distances in CFmMIMO systems result in Rician channel models with prominent line of sight (LoS) components between APs and UEs, motivating cooperation between APs for improved performance. 
    In this paper, we propose a cooperative minimum-mean-squared-error (MMSE)-based uplink CE approach where APs share their linearly compressed signals as \emph{fused} signals with other APs in the same cluster. The proposed approach is optimal, \ie its performance is equivalent to that of the centralized CE approach, where APs share their uncompressed raw signals.
    Notably, this optimality is achieved in one shot; that is, given the required correlation matrices, the optimal fusion filters and estimators are derived non-iteratively. Consequently, the proposed approach guarantees lower communication overhead for cooperative CE compared to the centralized approach.
    Numerical experiments corroborate the superior performance of the proposed cooperative CE approaches in terms of CE accuracy and convergence rate.
\end{abstract}

\begin{keywords}
    Cell-free massive multi-output-multi-input communication networks, channel estimation, distributed adaptive node-specific signal estimation
\end{keywords}

%% file: TeX/Text/introduction.tex
Massive multi-input-multi-output (mMIMO) communication networks \cite{bjornson_massive_2017,sanguinetti2019toward} have gained considerable attention over the years and have been deployed in the current 5G standard \cite{parkvall2017nr}. This technology relies on beamforming the signal in the downlink through multiple co-located antennas at an access point (AP) toward a specific user equipment (UE), enabling simultaneous transmission of data to multiple UEs by reusing common time and frequency resources. In the uplink, multiple co-located antennas at an AP are also used to estimate the signal from a specific UE by canceling interference from other UEs. APs are typically located at the center of non-overlapping dedicated regions called cells. Although a \emph{cellular} approach guarantees a scalable communication network with co-located computation at each cell, it suffers from a non-uniform quality of service (QoS) at the cell edges, where UEs are far from their serving AP while receiving considerable interference from APs in the neighboring cells \cite{jungnickel2014role}.

The non-uniform QoS in cellular mMIMO systems has motivated the development of network MIMO \cite{venkatesan2007network} and subsequently cell-free mMIMO (CFmMIMO) \cite{ngo_cell-free_2017,nayebi2017precoding} communication networks, where APs are distributed across the area, eliminating the concept of cells and thereby addressing the inferior performance at the cell edges. In the initial introduction of CFmMIMO, it was assumed that APs would have network-wide channel state information (CSI) and that each AP would serve all UEs in the network. While these assumptions are theoretically advantageous, they are challenging to implement in practice due to the significant computational complexity and fronthaul communication overhead between APs. 

To make CFmMIMO communication networks more practical, a \emph{user-centric} variant \cite{buzzi2017cell,bursalioglu2018fog} has been introduced, where each UE selects a cluster of APs with the best channel conditions (\eg the closest APs) and is served only by these APs rather than by all APs. In this approach, APs that are far from a UE are excluded to reduce communication and computational overhead without significant performance loss, making the CFmMIMO communication network scalable \cite{bjornson_scalable_2020}. Additionally, the clusters are dynamic and overlapping, meaning that as UEs relocate, clusters may change, and each AP may belong to multiple clusters, serving multiple UEs.

A key step in the aforementioned mMIMO and CFmMIMO communication networks is the estimation of the channels between a UE and the corresponding APs, which is a prerequisite for subsequent phases such as beamforming and decoding. In mMIMO systems, channel estimation (CE) typically occurs at each AP for the co-located antennas. In contrast, in CFmMIMO systems, CE can be performed locally at each AP or in a centralized manner within clusters, where APs in a cluster send their received uplink pilot signals via fronthaul links to a centralized processing unit (CPU) as a master AP in the cluster serving the UE for centralized CE. Among various CE approaches, a standard method is the minimum-mean-squared-error (MMSE)-based CE \cite{bjornson_scalable_2020,bjornson2019making}. This approach requires perfect knowledge of the channel correlation matrix, which can be estimated by various channel correlation estimation methods \cite{xie_overview_2016,bjornson2016massive,haghighatshoar2017massive,neumann2018covariance}. These methods typically require additional pilot overhead or complete knowledge of pilot assignments for all UEs in the network. A recent approach \cite{van_rompaey_gevd-based_2022} eliminates these requirements by using two data-driven correlation matrices: one representing the correlation of received uplink pilot signals and the other representing the correlation of corresponding despread signals. In this paper, we utilize this method to propose a \emph{cooperative} uplink CE within each AP cluster.

Existing CE approaches have typically been introduced in the mMIMO context, where CE is performed locally at each AP, and where the correlation between the channels of the co-located antennas is fully captured by various correlation estimation methods. In mMIMO communication networks, there is typically no correlation between channels of different APs, \ie the larger distances between APs result in uncorrelated channels, making cooperation unnecessary (cf. \cref{sec:channel_model} for more details on channel models). 
In contrast, in user-centric CFmMIMO communication networks, multiple APs in a cluster are closer to the corresponding UE, increasing the probability of line-of-sight (LoS) components between the UE and the APs, motivating cooperation for CE. The LoS component, along with the non-LoS (NLoS) component resulting from multi-path and scattering, increases the probability of experiencing a Rician channel model between the APs and the corresponding UE, rather than the Rayleigh channel model typically used in mMIMO systems.
Although the Rician channel model has been considered in various existing works in the CFmMIMO literature \cite{ozdogan_cell-free_2018,wang2020uplink,jin2020spectral,zhang2020noma,mukherjee_performance_2022,sun_bandwidth-efficient_2021,amadid_channel_2022}, the CE process is still conducted either locally at each AP or in a centralized manner, where APs share their raw signals with a CPU. A cooperative uplink CE approach has been introduced in \cite{van2024distributed}, where APs share their linearly compressed signals as \emph{fused} signals to limit the communication overhead between cooperating APs. However, this method is not optimal, \ie it does not achieve the performance of the centralized approach. Investigating optimal distributed signal estimation methods for cooperation between APs, efficient in both communication bandwidth and computational complexity, is of significant relevance and hence studied in this paper.

The distributed adaptive node-specific signal estimation (DANSE) algorithm for (wireless) sensor networks \cite{bertrand2010distributed}, along with its subsequent variants such as \cite{hassani2015gevd,szurley2016topology}, offers a relevant approach to establishing cooperation between various \emph{nodes} in a network. In its basic form \cite{bertrand2010distributed}, each node has one or more sensors that receive a signal that is captured by all the nodes in the network, thus sharing a common signal subspace plus noise. In DANSE, each node aims to estimate its desired node-specific signal from the received signal (\eg denoising the received signal). To achieve this, nodes share their \emph{fused} signals with other nodes, which can then benefit from these signals to estimate their desired node-specific signal. Deriving optimal fusion filters and estimators in DANSE is an iterative process, and it has been proven that the signal estimation performance converges to that of the centralized approach but with lower communication overhead. In a recent variant, called iteration-less DANSE (iDANSE) \cite{didier2024one}, signal estimation is optimal even if the signal subspaces of the desired node-specific signals are not fully overlapping, \ie node-specific signals are summations of \emph{global} signals correlated between nodes and \emph{local} signals uncorrelated between nodes, a scenario that matches with the Rician channel model with LoS (correlated) and NLoS (uncorrelated) components. Unlike DANSE, iDANSE achieves optimal performance without the need for an iterative procedure under certain assumptions. 

In this paper, iDANSE will be used for cooperative uplink CE in a CFmMIMO network. Specifically, a cooperative uplink CE approach is proposed in which the APs in a cluster cooperate for the CE for the corresponding UE. The estimation performance reaches that of the centralized approach, while instead of sharing uncompressed raw signals, the APs share fused signals, which reduces the communication bandwidth in the fronthaul links. Relevant numerical experiments are conducted to assess the performance of the proposed cooperative CE approach in terms of CE accuracy and convergence rate.

%% file: TeX/Text/nework_model.tex
We consider a CFmMIMO communication network consisting of $K$ single-antenna UEs and $L$ APs, each equipped with $N$ antennas. Similar to mMIMO communication networks, it is typically assumed that the number of antennas is larger than the number of UEs, \ie $LN \gg K$ \cite{kassam_review_2023,ngo_cell-free_2017}. To coordinate signal transmission and reception, APs are connected to CPUs through fronthaul links.

%% file: TeX/Text/user_centeric_model.tex
We assume a framework where each UE is served by a cluster of APs rather than all APs in the network. This framework guarantees a scalable realization of CFmMIMO, even when there are a large number of UEs in the network, \ie, $K \to \infty$ \cite{bjornson_scalable_2020}.
Define the cluster of APs serving UE $k$ as 
\begin{equation}
    \mathcal{A}_k \coloneqq \{\ell \mid \text{AP $\ell$ serves UE $k$}\}, ~\text{with}~ |\mathcal{A}_k| = L_k.
    \label{eq:Mk}
\end{equation}
The AP clustering can be either \emph{network-wide}, where APs are divided into fixed regional clusters, or \emph{user-centric}, where APs are divided into dynamic clusters for individual UEs. The user-centric arrangement provides a more uniform QoS throughout the area, hence in this paper, we use this arrangement for defining AP clusters $\mathcal{A}_k$ in \eqref{eq:Mk}. However, both arrangements can be addressed by the proposed methodologies in this paper.
With the user-centric arrangement, APs that receive the signal transmitted by UE $k$ with a high enough SINR form the cluster $\mathcal{A}_k$. Examples of user-centric AP clustering can be found in \cite{bjornson_scalable_2020}.
Consistently, the subset of UEs served by AP $\ell$ is defined as
\begin{equation}
    \mathcal{U}_\ell \coloneqq \{k \mid \text{UE $k$ is served by AP $\ell$}\}.
    \label{eq:Uk}
\end{equation}

%% file: TeX/Text/channel_model.tex
The channel between UE $k$ and the APs in cluster $\mathcal{A}_k$ is denoted by $\bm{h}^k \coloneqq [\bm{h}_{1,k}^T, \ldots, \bm{h}_{L_k,k}^T]^T \in \C^{NL_k}$, which is a concatenation of the individual channels $\bm{h}_{\ell,k} \in \C^N$ between AP $\ell$ in cluster $\mathcal{A}_k$ and UE $k$. The channels are assumed to remain constant during a single \emph{coherence block} with block index $b \in \{1,2,\dots\}$.
Unlike in cellular mMIMO communication networks, where the likelihood of LoS components is typically negligible due to blockage between the UEs and the centrally located AP in the corresponding cell, in CFmMIMO communication networks with distributed APs, the probability of having LoS components is significant due to the shorter AP-UE distances. This probability has been quantified in works such as \cite{mukherjee_performance_2022,ozdogan_cell-free_2018} as being inversely proportional to the AP-UE distances.

In the considered CFmMIMO communication network, the channel $\bm{h}^k$ in coherence block $b$ is a sample of a correlated Rician fading distribution as in \cite{ozdogan_cell-free_2018}:
\begin{align}
    \begin{split}
        \bm{h}^k[b] \sim \mathcal{N}_{\mathcal{C}}(\mathcirc{\bm{h}}^k, \bm{G}^k) \implies &\bm{h}^k[b] = \mathcirc{\bm{h}}^k + \dot{\bm{h}}^k[b] \in \C^{NL_k} \\
        \text{with}~ &\dot{\bm{h}}^k[b] \sim \mathcal{N}_{\mathcal{C}}(\bm 0, \bm{G}^k),
    \end{split}
    \label{eq:channel}
\end{align}
where the mean $\mathcirc{\bm{h}}^k \coloneqq [\mathcirc{\bm{h}}_{1,k}^T, \dots, \mathcirc{\bm{h}}_{L_k,k}^T]^T$ is deterministic and represents the LoS component, taking a non-zero value if the LoS component exists.
The NLoS component is represented by $\dot{\bm{h}}^k \coloneqq [\dot{\bm{h}}_{1,k}^T, \dots, \dot{\bm{h}}_{L_k,k}^T]^T$, with the small-scale fading modeled by a circularly symmetric complex normal distribution $\mathcal{N}_{\mathcal{C}}$ with spatial correlation matrix $\bm{G}^k \in \C^{NL_k \times NL_k}$. This matrix models large-scale fading effects such as path loss, shadowing, spatial channel correlation, and antenna gains. The Rician channel model in \eqref{eq:channel} is shown to be more accurate in CFmMIMO communication networks compared to the Rayleigh channel model \cite{mukherjee_performance_2022}.

The NLoS components $\dot{\bm{h}}_{\ell,k}$ are assumed independent across different APs due to the large spatial distribution of APs in the network. Therefore, 
\begin{align}
    \begin{split}
        \bm{R}^k \coloneqq &\E{\bm{h}^k[b] \bm{h}^{k^H}[b]} = \mathcirc{\bm{h}}^k \mathcirc{\bm{h}}^{k^H} + \bm{G}^k \in \C^{NL_k \times NL_k}\\
        \text{with}~~ &\bm{G}^k =~ \blkdiag\{\bm{G}_{1,k}, \dots, \bm{G}_{L_k,k}\},\\
        &\rank\{\bm{R}^k\} = \min\{1 + B L_k, N L_k\}, \\
        &B \coloneqq \rank\{\bm{G}_{\ell,k}\},
    \end{split}
    \label{eq:channel_cov}
\end{align}
where $B$ is considered common across all APs and UEs for simplicity.
The existence of LoS components motivates cooperation between APs within each cluster to enhance CE performance.
This cooperation will be discussed in \cref{sec:CE_cooperative}.

%% file: TeX/Text/DANSE.tex
In this section, we present the underlying signal model for CE within a more general framework for (wireless) sensor networks. Subsequently, the iDANSE algorithm \cite{didier2024one} for the fully connected topology is reviewed for cooperative MMSE-based signal estimation in this context. In \cref{sec:CE}, we employ iDANSE to propose a cooperative CE method.

\subsection{Sensor Network Signal Model}\label{sec:DANSE:signal_model}
We consider a sensor network with $L$ nodes each equipped with $N$ sensors. The noisy signal received by node $\ell$ with time index $b$ can be written as
\begin{align*}
    \bm{y}_\ell[b] 
    &= \bm{s}_\ell[b] + \bm{n}_\ell[b] \numberthis \label{eq:danse_signal_model}\\
    &= \mathcirc{\bm{s}}_\ell[b] + \dot{\bm{s}}_\ell[b] + \mathcirc{\bm{n}}_\ell[b] + \dot{\bm{n}}_\ell[b] \in\C^N,
\end{align*}
where $\bm{s}_\ell$ is the node-specific \emph{desired} signal to be estimated (denoised) and $\bm{n}_\ell$ is the noise signal at node $\ell$.
The node-specific desired signal contains the component $\mathcirc{\bm{s}}_\ell$, which includes contributions from $\mathcirc{S}$ \emph{latent} sources whose signals are captured by all the nodes; hence, we refer to these sources as \emph{global} latent sources.
The node-specific desired signal also contains the component $\dot{\bm{s}}_\ell$, which includes contributions from $\dot{S}_\ell$ \emph{latent} sources whose signals are captured only by node $\ell$; hence, we refer to these sources as \emph{local} latent sources. The node-specific signal $\bm{s}_\ell$ can then be written as
\begin{equation}
    \bm{s}_\ell[b] = \mathcirc{\bm{s}}_\ell[b] + \dot{\bm{s}}_\ell[b] = \sum_{k=1}^{\mathcirc{S}} \mathcirc{\bm{a}}_{\ell,k} \mathcirc{s}^{\rm lat}_k[b] + \sum_{k=1}^{\dot{S}_\ell} \dot{\bm{a}}_{\ell,k} \dot{s}^{\rm lat}_{\ell, k}[b].
\end{equation}
In the above expression, $\mathcirc{\bm{a}}_{\ell,k} \in \C^N$ and $\dot{\bm{a}}_{\ell,k} \in \C^N$ may be considered, for instance, as steering vectors that depend on the angle of arrival (AoA) of the latent sources. The signals $\mathcirc{s}^{\rm lat}_k \in \C$ and $\dot{s}^{\rm lat}_{\ell, k} \in \C$ from different latent sources are assumed to be independent, \ie $\E{\mathcirc{s}^{{\rm lat}}_k[b] \mathcirc{s}^{{\rm lat}^*}_q[b]} = 0$, $\E{\dot{s}^{{\rm lat}}_{\ell,k}[b] \dot{s}^{{\rm lat}^*}_{\ell,q}[b]} = 0$, and $\E{\mathcirc{s}^{{\rm lat}}_k[b] \dot{s}^{{\rm lat}^*}_{\ell,q}[b]} = 0$. 

Likewise, the noise signal $\bm{n}_\ell$ is decomposed into the component $\mathcirc{\bm{n}}_\ell$, which includes contributions from latent sources whose signals are captured by all the nodes (and are therefore correlated between the corresponding nodes), and the component $\dot{\bm{n}}_\ell$, which includes contributions from $\mathcirc{N}$ latent sources whose signals are captured only by node $\ell$. 

In node-specific signal estimation, each node $\ell$ aims to estimate $\bm{s}_{\ell}$---rather than the signals $\mathcirc{s}^{\rm lat}_k$ and $\dot{s}^{\rm lat}_{\ell, k}$, separately. The cooperation between nodes in node-specific signal estimation is motivated by the fact that the desired signals are correlated within an $\mathcirc{S}$-dimensional common signal subspace. In other words, $\rank \left\{\E{\bm{s}_{\ell}[b] \bm{s}_{j}^H[b]}\right\} = \rank\left\{\E{\mathcirc{\bm{s}}_{\ell}[b] \mathcirc{\bm{s}}_{j}^{H}[b]}\right\} = \rank\left\{\E{\mathcirc{\bm{s}}_{\ell}[b] \mathcirc{\bm{s}}_{\ell}^{H}[b]}\right\} = \mathcirc{S}, j \neq \ell$.

A centralized approach may be considered for estimation, where each node $\ell$ receives the \emph{$N$-dimensional} signals $\bm{y}_j \in \C^N$ from all other nodes $j\neq \ell$ and derives the following centralized MMSE estimator
\begin{equation}
    \bm W_{\ell} \coloneqq \argmin_{\bm W_{\ell} \in \C^{LN \times N}}{\E{\|\bm{s}_{\ell}[b] - \bm W_{\ell}^H \bm{y}[b]\|^2}},
    \label{eq:centralized_mmse}
\end{equation}
where $\bm{y} \coloneqq [\bm{y}_1^T,\dots,\bm{y}_L^T]^T\in\C^{LN}$ is the concatenation of the signal vectors from all nodes, and the node-specific signal is then estimated by $\hat{\bm{s}}_{\ell} = \bm W_{\ell}^H \bm{y}$. This estimation requires nodes to share $N$-dimensional signals, even though the underlying common signal subspace is $\mathcirc{S}$-dimensional due to the presence of $\mathcirc{S} < N$ global latent sources. 

As long as there are no local latent desired signal sources, i.e., $\dot{\bm{s}}_\ell = 0$ for all $\ell$, the DANSE algorithm, initially presented in \cite{bertrand2010distributed}, and its subsequent variants such as \cite{hassani2015gevd,szurley2016topology}, achieve the same performance as the centralized MMSE estimator \eqref{eq:centralized_mmse}. This is done by sharing \emph{fused} $J$-dimensional signals, with $J = \mathcirc{S}$, between nodes (as opposed to $N$-dimensional signals in the centralized approach) to iteratively derive a local MMSE estimator at each node, thereby saving communication bandwidth by a factor of $\nicefrac{N}{\mathcirc{S}}$.
However, these algorithms are not optimal (i.e., they do not achieve the performance of the centralized approach) in the presence of local latent sources, even if the fused signal dimension is increased to $\mathcirc{S} + \sum_\ell \dot{S}_{\ell}$. 
Recently, in \cite{didier2024one}, an iteration-less variant called iDANSE has been introduced, which is optimal by sharing fused signals of dimension  $J = \mathcirc{S} + \mathcirc{N}$ for the general signal model of \eqref{eq:danse_signal_model}. We briefly outline this algorithm in the next subsection and then discuss its iteration-less convergence.

\subsection{iDANSE Algorithm}
In the signal model \eqref{eq:danse_signal_model}, only the signals $\mathcirc{\bm{s}}_\ell$ and $\mathcirc{\bm{n}}_\ell$ originate from global latent sources that are captured by all the nodes. Hence, ideally nodes would share
\begin{equation}
    \mathcirc{\bm{y}}_\ell[b] \coloneqq \mathcirc{\bm{s}}_\ell[b] + \mathcirc{\bm{n}}_\ell[b],
\end{equation}
discarding local signal contributions, to achieve optimal performance. To achieve this, each node $\ell$ locally adopts the following MMSE estimator to estimate $J=\mathcirc{S} + \mathcirc{N}$ components of $\mathcirc{\bm{y}}_\ell$ from the received signal $\bm{y}_\ell$.
\begin{align}
    \begin{split}
        \mathcirc{\bm W}_\ell &\coloneqq \argmin_{\mathcirc{\bm W}_\ell \in \C^{N \times J}} \E{\|\bm{E}_J^T\mathcirc{\bm{y}}_\ell[b] - \mathcirc{\bm W}_\ell^{H} \bm{y}_\ell[b]\|^2}\\
        &= \bm R_{\bm{y}_\ell \bm{y}_\ell}^{-1} \bm R_{\bm{y}_\ell \mathcirc{\bm{y}}_\ell} \bm{E}_J,
    \end{split}
    \label{eq:idanse_local_W}
\end{align}
where $\bm R_{\bm{y}_\ell \bm{y}_\ell} \coloneqq \E{\bm y_\ell[b] \bm y_\ell^H[b]}$, $\bm R_{\bm{y}_\ell \mathcirc{\bm{y}}_\ell} \coloneqq \E{\bm{y}_\ell[b] \mathcirc{\bm{y}}_\ell^{H}[b]} = \E{\mathcirc{\bm{y}}_\ell[b] \mathcirc{\bm{y}}_\ell^{H}[b]} \eqqcolon \bm R_{\mathcirc{\bm{y}}_\ell \mathcirc{\bm{y}}_\ell}$, and $\bm{E}_J \coloneqq [\I_{J\times J}, \bm 0]^T \in \R^{N\times J}$ is a selection matrix that selects, without loss of generality, the first $J$ components of $\mathcirc{\bm{y}}_\ell$, assuming these are contributions from all global latent sources. While $\bm R_{\bm{y}_\ell \bm{y}_\ell}$ can be estimated from the observed $\bm y_\ell$, for the time being, it is assumed that $\bm R_{\mathcirc{\bm{y}}_\ell \mathcirc{\bm{y}}_\ell}$ can also be estimated.
With the MMSE estimator \eqref{eq:idanse_local_W}, each node estimates $\bm{z}_\ell[b] \coloneqq \mathcirc{\bm W}_\ell^{H} \bm{y}_\ell[b] \in \C^J$ and shares this estimate with all other nodes. Note that the estimate $\bm{z}_\ell$ is a fused version of the signal $\bm{y}_\ell$, with a fusion ratio of $\nicefrac{N}{J} > 1$, assuming $J < N$; hence, we call $\mathcirc{\bm W}_\ell$ the fusion filter.

\begin{algorithm}[t]
    \input{TeX/Alg/iDANSE.tex}
    \caption{iDANSE algorithm}
    \label{alg:idanse}
\end{algorithm}
Each node $\ell$ then receives the fused signals from the other $L-1$ nodes and concatenates these with its own signal $\bm{y}_\ell$ to form the \emph{observation} signal as follows:
\begin{equation}
    \tilde{\bm{y}}_\ell \coloneqq [\bm{y}_\ell^T,\bm{z}_{-\ell}^T]^T \in \C^{P}, ~\text{with}~ P \coloneqq N + (L - 1)J
    \label{eq:observation_vec}
\end{equation}
where $\bm{z}_{-\ell} \coloneqq [\bm{z}_{1}^T, \dots, \bm{z}_{\ell-1}^T, \bm{z}_{\ell+1}^T, \dots, \bm{z}_{L}^T]^T \in \C^{(L - 1)J}$.
Finally, the MMSE estimator of the desired node-specific signal $\bm{s}_\ell$ at each node $\ell$ can be derived by
\begin{align}
    \begin{split}
        \tilde{\bm{W}}_\ell &\coloneqq \argmin_{\tilde{\bm{W}}_\ell \in \C^{P \times N}} \E{\|\bm{s}_\ell[b] - \tilde{\bm{W}}_\ell^{H} \tilde{\bm{y}}_\ell[b]\|^2}\\
        &= \tilde{\bm R}_{\bm{y}_\ell \bm{y}_\ell}^{-1} \bm R_{\tilde{\bm{y}}_\ell \bm{s}_\ell},
    \end{split}
    \label{eq:idanse_W}
\end{align}
where $\bm R_{\tilde{\bm{y}}_\ell \tilde{\bm{y}}_\ell} \coloneqq \E{\tilde{\bm{y}}_\ell[b] \tilde{\bm{y}}_\ell^H[b]}$ and $\bm R_{\tilde{\bm{y}}_\ell \bm{s}_\ell} \coloneqq \E{\tilde{\bm{y}}_\ell[b] \bm{s}_\ell^H[b]}$. Again, for the time being, it is assumed that $\bm R_{\tilde{\bm{y}}_\ell \bm{s}_\ell}$ can also be estimated. In \cite{didier2024one}, it has been shown that the estimation performance of the estimator in \eqref{eq:idanse_W} is optimal, \ie equal to that of the centralized approach.

The iDANSE algorithm is presented in \cref{alg:idanse}. 
In practice, the signals $\mathcirc{\bm{y}}_\ell$ and $\bm{n}_\ell$ may not be directly available at each node for the estimation of $\bm R_{\mathcirc{\bm{y}}_\ell \mathcirc{\bm{y}}_\ell}$, $\bm R_{\tilde{\bm{n}}_\ell \tilde{\bm{n}}_\ell}$, and $\bm R_{\tilde{\bm{y}}_\ell \bm{s}_\ell}$. In \cref{sec:CE_cooperative}, we will show that if the global signal $\mathcirc{\bm{y}}_\ell$ has higher power than the local signal $\dot{\bm{s}}_\ell + \dot{\bm{n}}_\ell$, the estimation of $\bm R_{\mathcirc{\bm{y}}_\ell \mathcirc{\bm{y}}_\ell}$ is possible without direct access to $\mathcirc{\bm{y}}_\ell$. Moreover, we will show that by despreading the appropriate signals, the estimation of $\bm R_{\tilde{\bm{y}}_\ell \bm{s}_\ell}$ is possible without access to $\bm{n}_\ell$. 

In iDANSE, correlation matrices are estimated using averaging over coherence blocks with an exponential weighting factor of $\beta \in [0,1)$. 
The updates of $\mathcirc{\bm W}_\ell$ and $\tilde{\bm{W}}_\ell$, following \cref{alg:step:2,alg:step:3,alg:step:6,alg:step:7} in the \emph{for loop}, serve solely to improve the correlation matrix estimates or to track their changes in the case of long-term stationarity through sample averaging.
Unlike DANSE, once the correlation matrices are known, iDANSE achieves optimality with no iterations.

%% file: TeX/Alg/iDANSE.tex
\begin{algorithmic}[1]
    
    \item[\algfont{Initialize}]
    $\bm R_{\mathcirc{\bm{y}}_\ell \mathcirc{\bm{y}}_\ell}[0] \in \C^{N\times N}, \bm R_{\tilde{\bm{y}}_\ell \tilde{\bm{y}}_\ell}[0] \in \C^{P\times P}, \bm R_{\tilde{\bm{n}}_\ell \tilde{\bm{n}}_\ell}[0] \in \C^{P\times P}$, where $P \coloneqq N + J(L - 1) $, $\beta\in [0,1)$, and the batch size $\tau > 0$
        
    \hspace{5pt}
    \item[\algfont{For $b=1,2,\dots$ nodes synchronously repeat:}]
    \State\label{alg:step:1}
    Sample the signals $\bm{y}_\ell[b,i], \bm{n}_\ell[b,i]$, and $\mathcirc{\bm{y}}_\ell[b,i] \coloneqq \mathcirc{\bm{s}}_\ell[b,i] + \mathcirc{\bm{n}}_\ell[b,i]$, where each is the $i$-th sample in the batch of size $\tau$ at node $\ell$ and iteration $b$
    \State\label{alg:step:2}
    Update the local correlation matrices: 
    \begin{talign*}
        \bm R_{\bm{y}_\ell \bm{y}_\ell}[b] &= \beta \bm R_{\bm{y}_\ell \bm{y}_\ell}[b-1] + (1-\beta) \frac{1}{\tau} \sum_{i=1}^\tau \bm{y}_\ell[b,i] \bm{y}^H_\ell[b,i]\\
        \bm R_{\mathcirc{\bm{y}}_\ell \mathcirc{\bm{y}}_\ell}[b] &= \beta \bm R_{\mathcirc{\bm{y}}_\ell \mathcirc{\bm{y}}_\ell}[b-1] + (1-\beta) \frac{1}{\tau} \sum_{i=1}^\tau \mathcirc{\bm{y}}_\ell[b,i] \mathcirc{\bm{y}}^H_\ell[b,i]
    \end{talign*}
    \State\label{alg:step:3}
    Update the fusion filter:
    \begin{equation*}
        \mathcirc{\bm W}_\ell[b] = \bm R_{\bm{y}_\ell \bm{y}_\ell}^{-1}[b] \bm R_{\mathcirc{\bm{y}}_\ell \mathcirc{\bm{y}}_\ell}[b] \bm{E}_J \in \C^{N\times J}
    \end{equation*}
    where $\bm{E}_J \coloneqq [\I_{J\times J}, \bm 0]^T \in \R^{N\times J}$
    \State\label{alg:step:4}
    Compute and broadcast the following fused signals to all other nodes:
    \begin{align*}
        \bm{z}_\ell[b,i] &\coloneqq \mathcirc{\bm W}_\ell^{H}[b] \bm{y}_\ell[b,i] \in \C^J \\ 
        \bm{q}_\ell[b,i] &\coloneqq \mathcirc{\bm W}_\ell^{H}[b] \bm{n}_\ell[b,i] \in \C^J
    \end{align*}
    \State\label{alg:step:5}
    Form the observation signals by the received fused signals:
    \begin{align*}
        \tilde{\bm{y}}_\ell[b,i] &\coloneqq [\bm{y}_\ell^T[b,i],\bm{z}_{-\ell}^T[b,i]]^T \in \C^P, \\ 
        \tilde{\bm{n}}_\ell[b,i] &\coloneqq [\bm{n}_\ell^T[b,i],\bm{q}_{-\ell}^T[b,i]]^T \in \C^P,
    \end{align*}
    where $\bm{z}_{-\ell} \coloneqq [\bm{z}_{1}^T, \dots, \bm{z}_{\ell-1}^T, \bm{z}_{\ell+1}^T, \dots, \bm{z}_{L}^T]^T$ and $\bm{q}_{-\ell} \coloneqq [\bm{q}_{1}^T, \dots, \bm{q}_{\ell-1}^T, \bm{q}_{\ell+1}^T, \dots, \bm{q}_{L}^T]^T$
    \State\label{alg:step:6}
    Update the iDANSE correlation matrices: 
    \begin{talign*}
        \bm R_{\tilde{\bm{y}}_\ell \tilde{\bm{y}}_\ell}[b] &= \beta \tilde{\bm R}_{\bm{y}_\ell \bm{y}_\ell}[b-1] + (1-\beta) \frac{1}{\tau} \sum_{i=1}^\tau \tilde{\bm{y}}_\ell[b,i] \tilde{\bm{y}}_\ell^H[b,i]\\
        \bm R_{\tilde{\bm{n}}_\ell \tilde{\bm{n}}_\ell}[b] &= \beta \tilde{\bm R}_{\bm{n}_\ell \bm{n}_\ell}[b-1] + (1-\beta) \frac{1}{\tau} \sum_{i=1}^\tau \tilde{\bm{n}}_\ell[b,i] \tilde{\bm{n}}_\ell^H[b,i]\\
        \bm R_{\tilde{\bm{y}}_\ell \bm{s}_\ell}[b] &= (\bm R_{\tilde{\bm{y}}_\ell \tilde{\bm{y}}_\ell}[b] - \bm R_{\tilde{\bm{n}}_\ell \tilde{\bm{n}}_\ell}[b]) \tilde{\bm{E}}_N
    \end{talign*}
    where $\tilde{\bm{E}}_N \coloneqq [\I_{N\times N}, \bm 0]^T \in \R^{P\times N}$
    \State\label{alg:step:7}\label{alg:step:lambda_update}
    Update the iDANSE MMSE estimator:
    \begin{equation*}
        \tilde{\bm{W}}_\ell[b] = \bm R_{\tilde{\bm{y}}_\ell \tilde{\bm{y}}_\ell}^{-1}[b] \bm R_{\tilde{\bm{y}}_\ell \bm{s}_\ell}[b] \in \C^{P\times N}
    \end{equation*}
    \State\label{alg:step:8}
    Estimate (denoise) the desired node-specific signal:
    \begin{equation*}
        \hat{\bm{s}}_\ell[b,i] = \tilde{\bm{W}}_\ell^{H}[b] \tilde{\bm{y}}_\ell[b,i]
    \end{equation*}

\end{algorithmic}

%% file: TeX/Text/channel_estimation.tex
Considering the Rician channel model \eqref{eq:channel} for CFmMIMO communication networks, the LoS component of different APs motivates cooperation between APs within a cluster for CE, leading to improved performance compared to local CEs.
In this section, we first present a local CE approach in \cref{sec:CE_local}. Then, we present the proposed cooperative CE approach in \cref{sec:CE_cooperative}.

\subsection{Local Uplink Channel Estimation}\label{sec:CE_local}
    \input{TeX/Text/channel_estimation_local.tex}

\subsection{Cooperative Uplink Channel Estimation}\label{sec:CE_cooperative}
    \input{TeX/Text/channel_estimation_cooperative.tex}

%% file: TeX/Text/channel_estimation_local.tex
\begin{algorithm}[t]
    \input{TeX/Alg/CE_local.tex}
    \caption{Local Channel Estimation}
    \label{alg:CE_local}
\end{algorithm}
Existing CE approaches in mMIMO communication networks can also be extended to CFmMIMO communication networks \cite{bjornson_scalable_2020,bjornson_massive_2017}. However, due to the user-centric and cell-free nature, avoiding pilot contamination, where different UEs use the same pilot signal, becomes more challenging. In \cite{van_rompaey_gevd-based_2022}, a CE method is proposed in which pilots are randomly assigned to UEs in each channel coherence block. The proposed method in \cite{van_rompaey_gevd-based_2022} focuses on CE in mMIMO communication networks, where the channels between multiple UEs and a single AP do not have LoS components. Therefore, this subsection presents a slightly modified procedure based on \cite{van_rompaey_gevd-based_2022}, making it suitable for the case where channel LoS components are present along with the NLoS components\footnote{The modified procedure here also relaxes the assumption of independent LoS components unlike \cite{van2024distributed}, where additional phases $\varphi_k$ are considered for the LoS components, \ie $\mathcirc{\bm h}^k e^{j\varphi_k}$ with independent phases for different UEs.}.

Consider that each channel coherence block (with block index $b$) is divided into $\tau_c = \tau_p + \tau_d$ sample instants (with sample index $i$), where $\tau_p$ sample instants are allocated for pilot-based uplink CE, allowing for $\tau_p$ mutually orthogonal pilot sequences, followed by $\tau_d$ sample instants for data transmission \cite{ngo_cell-free_2017,bjornson_new_2019}.
Note that the length of each channel coherence block depends on the carrier frequency and external factors such as the propagation environment and UE mobility. In each coherence block, UE $k$ randomly selects a pilot sequence $s^p[i], i\in\{1,\dots,\tau_p\}$, which is the $p$th pilot sequence out of $\tau_p$ orthogonal pilot sequences, where $\sum_{i=1}^{\tau_p} (s^{p}[i])^2 = \tau_p$. In addition, the UE either flips the sequence with a probability of 0.5 or leaves it unchanged. We denote the transmitted pilot by $s_k^p[b,i] \coloneqq \gamma_k[b] s^p[i]$, with the random variable $\gamma_k[b] \in \{+1,-1\}$. The received signal at AP $\ell \in \mathcal{A}_k$ and coherence block $b$ can then be written as follows:
\begin{align}
    \begin{split}
        \bm{y}_\ell^p[b,i] \coloneqq \sum_{k\in\mathcal{U}_\ell} \bm{h}_{\ell,k}[b] s_k^p[b,i] + \bm{n}_\ell[b,i] \in \C^N, \quad i\in[\tau_p],
    \end{split}
    \label{eq:received_sig}
\end{align}
where $\bm{h}_{\ell,k}$ represents the channel between AP $\ell$ and UE $k$, and $\bm{n}_\ell$ indicates the independent noise at AP $\ell$. In \eqref{eq:received_sig}, AP $\ell$ receives signals transmitted by the UEs in the set $\mathcal{U}_\ell$ as defined in \eqref{eq:Uk}, which are served by AP $\ell$ based on a high received SINR at AP $\ell$. The other UEs that do not belong to $\mathcal{U}_\ell$ have negligible received SINR at AP $\ell$ and are, therefore, not included.

It is noteworthy that in this CE approach, the pilot selection occurs randomly at each coherence block, which differs from the procedures commonly considered in the literature, \eg \cite{bjornson_scalable_2020}. In most existing works, pilots are typically assigned to minimize pilot contamination. However, in scenarios where the number of UEs served by an AP cluster exceeds the number of pilot sequences $\tau_p$, pilot contamination may become significant, necessitating techniques for pilot decontamination. An advantage of the CE approach presented here is that it does not require such considerations and decontamination techniques, as will become clear.

It is assumed that if UE $k \in \mathcal{U}_\ell$ is served by AP $\ell$, then, AP $\ell$ knows the pilot $s_k^p$ chosen by UE $k \in \mathcal{U}_\ell$ in each coherence block and, hence, can take the inner product between its received signal $\bm{y}_\ell^p$ and the pilot $s_k^p$ to obtain sufficient statistics (to \emph{despread} the received signal) for the estimation of $\bm{h}_{\ell,k}$. This results in
\begin{align*}
        &\bm{y}_{\ell,k}^{p,\rm desp}[b] \coloneqq \sum_{i=1}^{\tau_p} s_k^{p}[b,i] \bm{y}_\ell^p[b,i] \in \C^N\numberthis \label{eq:despread_sig}\\
        &= \tau_p \bm{h}_{\ell,k}[b] +  \tau_p\sum_{q\in\mathcal{U}_\ell,q\neq k} \bm{h}_{\ell,q}[b]\delta_q[b] + \sum_{i=1}^{\tau_p} s_k^{p}[b,i] \bm{n}_\ell[b,i],
\end{align*}
where, due to the randomness of the pilot selection and pilot flipping, $\delta_q[b]$ is either $0$ if UE $q$ picks a pilot different from $s^p$ (due to the pilot orthogonality), which happens with probability $1 - \frac{1}{\tau_p}$, or $\delta_q[b] \in \{-1, 1\}$ if UE $q$ picks the same pilot $s^p$ with either $\gamma_q=-1$ or $\gamma_q=1$, which happens with probability $\frac{1}{\tau_p}$. Therefore, $\delta_q[b]$ is a random variable with $\E{\delta_q[b]} = 0$ and $\E{\delta_q^2[b]} = \frac{1}{\tau_p}$. 

The aim is now to obtain an optimal MMSE-based channel estimator to estimate $\bm{h}_{\ell,k}[b]$ from the despread signal $\bm{y}_{\ell,k}^{p,\rm desp}$ as follows
\begin{align*}
    \bm{W}_{\ell,k} &\coloneqq \argmin_{\bm{W}_{\ell,k} \in \C^{N\times N}}\E{\|\bm{h}_{\ell,k}[b] - \bm{W}_{\ell,k}^H \bm{y}_{\ell,k}^{p,\rm desp}[b]\|^2} \numberthis \label{eq:channel_estimator}\\
    &= \tau_p \left(\bm{R}^{{p, \rm desp}}_{\ell,k}\right)^{-1} \bm{R}_{\ell,k} \quad \forall k,
\end{align*}
where the correlation matrices are defined by $\bm{R}^{p,\rm desp}_{\ell,k} \coloneqq \E{\bm{y}_{\ell,k}^{p,\rm desp}[b] \bm{y}_{\ell,k}^{p,{\rm{desp}}, H}[b]}$ and $\bm{R}_{\ell,k} \coloneqq \E{\bm{h}_{\ell,k}[b]\bm{h}_{\ell,k}^H[b]}$. To calculate the unknown correlation $\bm{R}_{\ell,k}$, consider the following
\begin{align*}
    \bm{R}_\ell^p &\coloneqq \E{\sum_{i=1}^{\tau_p} \bm{y}_\ell^p[b,i] \bm{y}_\ell^p[b,i]^H} \\
    & \quad\quad = \tau_p \bm{R}_{\ell,k} + \tau_p \sum_{q\in\mathcal{U}_\ell,q\neq k} \bm{R}_{\ell,q} + \tau_p\bm{R}_{n_\ell n_\ell}, \numberthis \label{eq:despreading_covs} \\
    \bm{R}^{p,\rm desp}_{\ell,k} &\coloneqq \E{\bm{y}_{\ell,k}^{p,\rm desp}[b] \bm{y}_{\ell,k}^{p,{\rm{desp}}, H}[b]} \\
    & \quad\quad = \tau_p^2 \bm{R}_{\ell,k} + \tau_p \sum_{q\in\mathcal{U}_\ell,q\neq k} \bm{R}_{\ell,q} + \tau_p\bm{R}_{n_\ell n_\ell}, \numberthis \label{eq:despreading_covs_desp}
\end{align*}
with $\bm{R}_{n_\ell n_\ell} \coloneqq \E{\bm{n}_\ell[b,i]\bm{n}_\ell^H[b,i]}$. By despreading the received signal with the pilot $s_k^p$, \ie by calculating the inner product in \eqref{eq:despread_sig}, AP $\ell$ gives priority to UE $k$ over other UEs. 
Comparing the two correlation matrices in \eqref{eq:despreading_covs} and \eqref{eq:despreading_covs_desp}, it is evident that UE $k$ is prioritized by AP $\ell$ in the despread signal $\bm{y}_{\ell,k}^{p,\rm desp}$, as the correlation matrix of the channel between AP $\ell$ and UE $k$, \ie $\bm{R}_{\ell,k}$, is boosted by a factor of $\tau_p$ in $\bm{R}^{p,\rm desp}_{\ell,k}$.

With \eqref{eq:despreading_covs} and \eqref{eq:despreading_covs_desp}, the channel correlation matrix $\bm{R}_{\ell,k}$ can be expressed as
\begin{equation}
    \bm{R}_{\ell,k} = \frac{1}{\tau_p^2 - \tau_p} (\bm{R}^{p,\rm desp}_{\ell,k} - \bm{R}_\ell^p).
    \label{eq:estimated_cov}
\end{equation}
Alternatively, $\bm{R}_{\ell,k}$ can also be estimated using generalized eigenvalue decomposition (GEVD)-based low-rank approximation techniques, \ie by exploiting the GEVD of the matrix pencil $\{\bm{R}^{p,\rm desp}_{\ell,k}, \bm{R}_\ell^p\}$ \cite{van_rompaey_gevd-based_2022}.

After deriving the channel estimator $\bm{W}_{\ell,k}$ in \eqref{eq:channel_estimator} the estimated channel is obtained by
\begin{equation}
    \hat{\bm{h}}_{\ell,k}[b] \coloneqq \bm{W}_{\ell,k}^H \bm{y}_{\ell,k}^{p,\rm desp}[b] \quad \forall \ell, \forall k \in \mathcal{U}_\ell, \forall b.
\end{equation}
The presented local CE approach is summarized in \cref{alg:CE_local}, where the correlation matrices are estimated by sample averaging over coherence blocks, with an exponential weighting factor $\beta$.

%% file: TeX/Alg/CE_local.tex
\begin{algorithmic}[1]
    
    \item[\algfont{Initialize}]
    Orthogonal pilot sequences $s^p, p\in \{1,...,\tau_p\}$, UE subsets $\mathcal{U}_\ell$, defined in \eqref{eq:Uk} for each AP $\ell\in\{1,...,L\}$, $\bm{R}_\ell^{p}[0] \in \C^{N\times N}$, $\bm{R}^{p,\rm desp}_{\ell,k}[0] \in \C^{N\times N}, \forall \ell,k$, and $\beta \in [0,1)$
        
    \hspace{5pt}
    \item[\algfont{For each coherence block $b=1,2,\dots$:}]
    \item[~~\algfont{For each AP $\ell \in \{1,...,L\}$:}]
    \State\label{alg:local:1}
    Sample the received signal $\bm{y}_\ell^{p}[b,i], i\in\{1,...,\tau_p\}$, defined in \eqref{eq:received_sig}, during the CE step of coherence block $b$
    \State\label{alg:local:2}
    Update the local correlation matrix:
    \begin{talign*}
        \bm{R}_\ell^{p}[b] &= \beta \bm{R}_\ell^{p}[b-1] + (1-\beta) \sum_{i=1}^{\tau_p} \bm{y}_\ell^{p}[b,i] \bm{y}_\ell^{p^H}[b,i]
    \end{talign*}
    \State\label{alg:local:3}
    {\algfont{For each UE $k\in\mathcal{U}_\ell$:}}
    \begin{algsubstates}
        \item\label{alg:local:3:a} Derive the despread signal, $\bm{y}_{\ell,k}^{p,\rm desp}[b]$, by \eqref{eq:despread_sig}
        \item\label{alg:local:3:b} Update the local correlation matrices: 
        \begin{talign*}
            \bm{R}^{p,\rm desp}_{\ell,k}[b] &= \beta \bm{R}^{p,\rm desp}_{\ell,k}[b-1] \\
            &+ (1-\beta) \bm{y}_{\ell,k}^{p,\rm desp}[b] \bm{y}_{\ell,k}^{p,{\rm{desp}}^{H}}[b]\\
            \bm{R}_{\ell,k}[b] &= \frac{1}{\tau_p^2 - \tau_p} (\bm{R}^{p,\rm desp}_{\ell,k}[b] - \bm{R}_\ell^{p}[b])
        \end{talign*}
        \item\label{alg:local:3:c} Update the local MMSE CE estimators:
        \begin{equation*}
            \bm{W}_{\ell,k}[b] = \tau_p \left(\bm{R}^{p, \rm desp}_{\ell,k}[b]\right)^{-1} \bm{R}_{\ell,k}[b] \in \C^{N\times N}\quad \forall k,
        \end{equation*}
        \item\label{alg:local:3:d} Estimate the local channel $\bm{h}_{\ell,k}$:
        \begin{equation*}
            \hat{\bm{h}}_{\ell,k}[b] \coloneqq \bm{W}_{\ell,k}^{H}[b] \bm{y}_{\ell,k}^{p,\rm desp}[b] 
        \end{equation*}
    \end{algsubstates}        
\end{algorithmic}

%% file: TeX/Text/channel_estimation_cooperative.tex
In this subsection, we present the main contribution of the paper, \ie an algorithm where APs of each cluster $\mathcal{A}_k$, defined in \eqref{eq:Mk}, cooperate in the CE for UE $k$. 
We follow the same methodology presented in \cref{sec:CE_local} for CE, while the cooperation is governed by the iDANSE algorithm, outlined in \cref{sec:DANSE}. The proposed cooperative CE approach is summarized in \cref{alg:CE_cooperative}.

In \cref{alg:CE_cooperative}, each AP $\ell$ knows the set of UEs $\mathcal{U}_\ell$ it serves and the APs $j$ with which it cooperates for CE for each UE $k \in \mathcal{U}_\ell$. The clusters can be formed following \cref{sec:AP_assignment}.
For a cooperating pair of APs $\ell$ and $j$, the received signal at AP $\ell \in \mathcal{A}_k$ in coherence block $b$, defined in \eqref{eq:received_sig}, contains the same components as the signal model in \eqref{eq:danse_signal_model}, and is therefore compatible with the iDANSE algorithm, \ie
\begin{align*}
        \bm{y}_\ell^p[b,i] 
        &= \underbrace{\sum_{k\in\mathcal{U}_\ell \cap \mathcal{U}_j} \mathcirc{\bm{h}}_{\ell,k}[b] s_k^p[b,i]}_{\mathcirc{\bm{s}}_\ell+\mathcirc{\bm{n}}_\ell} \\
        &+ \underbrace{\sum_{k\in\mathcal{U}_\ell \setminus \mathcal{U}_j} \mathcirc{\bm{h}}_{\ell,k}[b] s_k^p[b,i] + \sum_{k\in\mathcal{U}_\ell} \dot{\bm{h}}_{\ell,k}[b] s_k^p[b,i]}_{\dot{\bm{s}}_\ell} \numberthis \label{eq:equivalence}\\
        &+ \underbrace{\bm{n}_\ell[b,i] }_{\dot{\bm{n}}_\ell}\in \C^N, \quad i\in[\tau_p].
\end{align*}
The equivalence of the signal models in \eqref{eq:danse_signal_model} and \eqref{eq:equivalence} motivates cooperation between APs in $\mathcal{A}_k$ for CE, \ie adopting iDANSE for each UE $k$. 

APs follow \cref{alg:proposed:1} to \cref{alg:proposed:2:d} of \cref{alg:CE_cooperative} to derive their fusion filters for each AP $j$ with which AP $\ell$ cooperates, given the received signal $\bm{y}_{\ell}^{p}$. The fusion filters are estimators for the correlated part of the signal $\bm{y}_{\ell}^{p}$, \ie
\begin{equation}
    \mathcirc{\bm{y}}_{\ell \to j}^p[b,i] 
    \coloneqq \sum_{k\in\mathcal{U}_\ell \cap \mathcal{U}_j} \mathcirc{\bm{h}}_{\ell,k}[b] s_k^p[b,i], \quad \ell,j\in[L].
    \label{eq:ylk} 
\end{equation}
These steps are equivalent to \cref{alg:step:1,alg:step:2,alg:step:3,alg:step:4} in \cref{alg:idanse}. Recall that according to the iDANSE terminology, $\mathcirc{\bm{y}}_{\ell\to j}^p \equiv \mathcirc{\bm{s}}_\ell+\mathcirc{\bm{n}}_\ell$ is the signal captured by node $\ell$ and some other node $j$. Consequently, AP $\ell$ only shares $\mathcirc{\bm{y}}_{\ell\to j}^p$ with AP $j$, discarding local signal contributions. Note that at the later steps in \cref{alg:CE_cooperative}, depending on the UE for which the CE is being performed, $\mathcirc{\bm{s}}_\ell \equiv \mathcirc{\bm{h}}_{\ell,k}[b] s_k^p[b,i]$ and $\mathcirc{\bm{n}}_\ell \equiv \sum_{q\in \mathcal{U}_\ell \cap \mathcal{U}_j \setminus k} \mathcirc{\bm{h}}_{\ell,q}[b] s_q^p[b,i]$. According to iDANSE, the optimality is guaranteed with sharing $J_{\ell\to j} \coloneqq |\mathcal{U}_\ell \cap \mathcal{U}_j|$-dimensional signals between nodes (refer to the signal dimension $J$ in \eqref{eq:idanse_local_W}).
Therefore, definig $\bm e_{\ell\to j} \coloneqq [\I_{J_{\ell\to j}\times J_{\ell\to j}},\bm 0]^T \in \R^{N\times J_{\ell\to j}}$ as a selection matrix, the fusion filter can be derived as 
\begin{align*}
    \mathcirc{\bm{W}}_{\ell\to j} &\coloneqq \argmin_{\mathcirc{\bm{W}}_{\ell\to j}}\E{\|\bm e_k^T\mathcirc{ \bm{y}}_{\ell\to j}^{p}[b,i] - \mathcirc{\bm{W}}_{\ell\to j}^H \bm{y}_{\ell}^{p}[b,i]\|^2} \numberthis \label{eq:LOS_estimator}\\
    &= \bm{R}^{{p}^{-1}}_{\ell} \bm{R}_{\mathcirc{\bm{y}}_{\ell\to j}^{p} \mathcirc{\bm{y}}_{\ell\to j}^{p}} \bm e_k \in \C^{N\times J_{\ell\to j}}, 
\end{align*}
where 
\begin{align}
    \bm{R}_{\mathcirc{\bm{y}}_{\ell\to j}^{p} \mathcirc{\bm{y}}_{\ell\to j}^{p}} \coloneqq \E{\mathcirc{ \bm{y}}_{\ell\to j}^{p}[b,i] \mathcirc{ \bm{y}}_{\ell\to j}^{p^H}[b,i]} = \tau_p\sum_{k\in \mathcal{U}_{\ell} \cap \mathcal{U}_{j}} \mathcirc{\bm{R}}_{\ell,k},
    \label{eq:Ryy}
\end{align}
with $\mathcirc{\bm{R}}_{\ell,k} \coloneqq \mathcirc{\bm{h}}_{\ell,k} \mathcirc{\bm{h}}_{\ell,k}^H$.
Considering the channel model in \eqref{eq:channel} together with \eqref{eq:estimated_cov}, and the fact that for the typical AP-UE distances in CFmMIMO communication networks the LoS component has more power compared to the NLoS component (the quantitative comparisons are provided in the numerical experiments in \cref{sec:sim}), the matrices $\mathcirc{\bm{R}}_{\ell,k}$ can be derived by rank-1 approximations of $\bm{R}_{\ell,k}$, \ie $\mathcirc{\bm{R}}_{\ell,k} = (\bm{R}_{\ell,k})_{\rm rank=1} = \frac{1}{\tau_p^2 - \tau_p}(\bm{R}^{p,\rm desp}_{\ell,k} - \bm{R}_\ell^{p})_{\rm rank=1}$ with $k\in\mathcal{U}_\ell$.
These approximations can be performed using either EVD or GEVD \cite{hassani2015gevd,van_rompaey_gevd-based_2022}, where the largest eigenvalue corresponding to the LoS component is selected for the rank-1 approximation. 

\begin{figure}[t]    
    \centering    
    \includegraphics[width=\columnwidth]{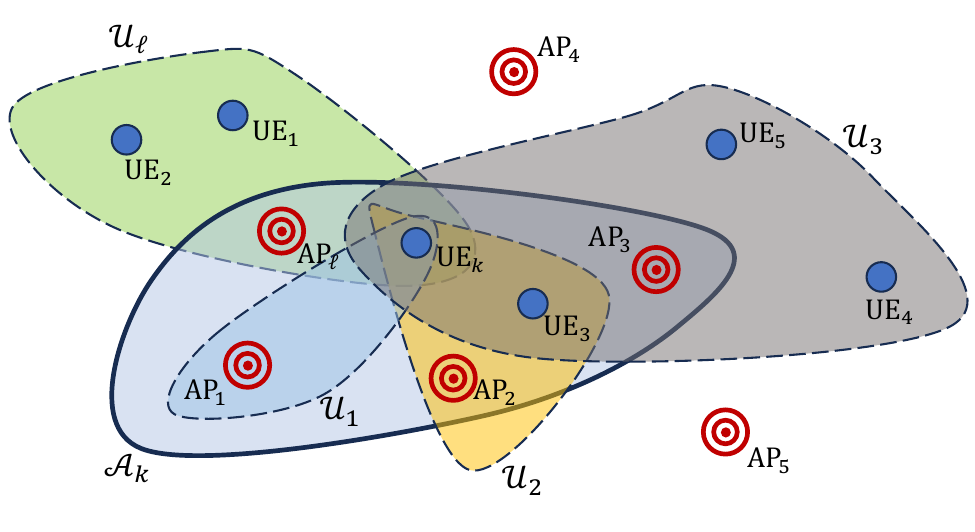}    
    \caption{An AP cluster $\mathcal{A}_k=\{AP_1,AP_2,AP_3,AP_\ell\}$ serving UE $k$ and cooperating for the estimation of channels $\bm{h}_{j,k}, j\in\{1,2,3,\ell\}$. UE subsets $\mathcal{U}_j$ served by the APs in $\mathcal{A}_k$ are also depicted. Following an AP clustering algorithm, APs know the involved clusters and relevant parameters for cooperative CE in $\mathcal{A}_k$, such as $J_{\ell\to j} \coloneqq |\mathcal{U}_\ell \cap \mathcal{U}_j|$. Some examples are $J_{2\to3}=J_{3\to2}=2$ and $J_{2\to1}=J_{1\to2}=J_{2\to\ell}=J_{\ell\to2}=1$.}    
    \label{fig:sets}
\end{figure}
Following \cref{alg:proposed:2:e} in \cref{alg:CE_cooperative}, each AP concatenates the received fused signals with its own $N$-dimensional signal $\bm{y}_\ell^{p}[b,i]$ to form the $P_{\ell,k}$-dimensional observation signal $\tilde{\bm{y}}_{\ell,k}^{p}$ with $P_{\ell,k} \coloneqq N + \sum_{q\in\mathcal{A}_k\setminus\ell} J_{q\to \ell}$. After despreading the fused signal at \cref{alg:proposed:2:f1}, the corresponding correlation matrices are then estimated by sample-averaging with an exponential factor $\beta$. It is noted that the estimation of the channel correlation $\tilde{\bm{R}}_{\ell,k}$ follows the same reasoning in \cref{alg:local:3:b} of \cref{alg:CE_local} for local CE. Moreover, the mentioned steps are equivalent to \cref{alg:step:5,alg:step:6} of \cref{alg:idanse}.

At \cref{alg:proposed:2:g} of \cref{alg:CE_cooperative}, each AP derives its MMSE channel estimator, which solves the following problem: 
\begin{align*}
    \tilde{\bm{W}}_{\ell,k} &\coloneqq \argmin_{\tilde{\bm{W}}_{\ell,k} \in \C^{P_{\ell,k}\times N}}\E{\|\bm{h}_{\ell,k}[b] - \tilde{\bm{W}}_{\ell,k}^H \tilde{\bm{y}}_{\ell,k}^{p,\rm desp}[b]\|^2}\\
    &= \tau_p \left(\tilde{\bm{R}}_{\ell,k}^{p,{\rm desp}}\right)^{-1} \tilde{\bm{R}}_{\ell,k} \tilde{\bm{E}}_k,
\end{align*}
where $\tilde{\bm{E}}_k \coloneqq [\I_{N\times N}, \bm 0]^T \in \R^{ P_{\ell,k} \times N}$. This step follows the MMSE local estimator in \cref{alg:local:3:c} of \cref{alg:CE_local} and is equivalent to \cref{alg:step:7} of \cref{alg:idanse}.
Finally, at \cref{alg:proposed:2:h}, APs estimate their channel $\bm{h}_{\ell,k}$ to UE $k$, which is an equivalent step to the local CE in \cref{alg:local:3:d} of \cref{alg:CE_local} and \cref{alg:step:8} of \cref{alg:idanse}.

It is noted that, similar to iDANSE, given the required correlation matrices, the fusion filters $\mathcirc{\bm W}_{\ell\to j}$ and estimators $\tilde{\bm{W}}_{\ell,k}$ are derived in a non-iterative manner. Their updates, as outlined in \cref{alg:proposed:2:c,alg:proposed:2:g} within the for loops of \cref{alg:CE_cooperative}, serve solely to improve the correlation matrix estimates or to track their changes in the case of long-term stationarity through sample averaging. 
\begin{algorithm}[th]
    \input{TeX/Alg/CE_cooperative.tex}
    \caption{iDANSE-based Cooperative Channel Estimation}
    \label{alg:CE_cooperative}
\end{algorithm}

%% file: TeX/Alg/CE_cooperative.tex
\begin{algorithmic}[1]
    
    \item[\algfont{Initialize}]
    Orthogonal pilot sequences $s^p, p\in\{1,...,\tau_p\}$,
    clusters $\mathcal{A}_k$ with $|\mathcal{A}_k| = L_k$ and subsets $\mathcal{U}_\ell$, defined in \eqref{eq:Mk} and \eqref{eq:Uk}, for all $k$ and $\ell$, 
    $\mathcirc{\bm{R}}_{\ell,k}[0] \in \C^{N\times N}$, $\tilde{\bm{R}}_{\ell,k}^{p,{\rm desp}}[0] \in \C^{P_{\ell,k} \times P_{\ell,k}}, \tilde{\bm{R}}_{\ell,k}^{p}[0] \in \C^{P_{\ell,k} \times P_{\ell,k}}$, where $P_{\ell,k} \coloneqq N + \sum_{q\in\mathcal{A}_k\setminus\ell} J_{q\to \ell}$, and $\beta \in [0,1)$
        
    \hspace{5pt}
    \item[\algfont{For each coherence block $b=1,2,\dots$:}]
    \item[~~\algfont{For each AP $\ell = 1, \dots, L$:}]
    \State\label{alg:proposed:1}
    Sample the received signal $\bm{y}_\ell^{p}[b,i], i\in\{1,...,\tau_p\}$, defined in \eqref{eq:received_sig}, during the CE step of coherence block $b$
    \State\label{alg:proposed:1:1}
    {\algfont{For each AP $j$ with which AP $\ell$ cooperates\\ (\ie $\mathcal{U}_\ell \cap \mathcal{U}_j \neq \emptyset, ~j\in\{1,\dots,L\}$):}}
    \begin{algsubstates}
        \item\label{alg:proposed:2:c} Update the fusion filter:
        \begin{equation*}
            \mathcirc{\bm{W}}_{\ell\to j}[b] = (\bm{R}_{\ell}^{p}[b])^{-1} \bm{R}_{\mathcirc{\bm{y}}_{\ell\to j}^{p} \mathcirc{\bm{y}}_{\ell\to j}^{p}}[b] \bm e_{\ell\to j} \in \C^{N\times J_{\ell\to j}},
        \end{equation*}
        where $\bm{R}_{\mathcirc{\bm{y}}_{\ell\to j}^{p} \mathcirc{\bm{y}}_{\ell\to j}^{p}}[b]$ is derived in \eqref{eq:Ryy}, $\bm{R}_\ell^{p}[b]\in\C^{N\times N}$ is the upper-left block of the matrix $\tilde{\bm{R}}_{\ell,k}^{p}[b]$, and $\bm e_{\ell\to j} \coloneqq [\I_{J_{\ell\to j}\times J_{\ell\to j}},\bm 0]^T \in \R^{N\times J_{\ell\to j}}$ 
        \item\label{alg:proposed:2:d} Compute and transmit to AP $j$ the fused signal:
        \begin{align*}
            \bm{z}_{\ell\to j}[b,i] &\coloneqq \mathcirc{\bm{W}}_{\ell \to j}^{H}[b] \bm{y}_{\ell}^{p}[b,i] \in \C^{J_{\ell\to j}}, \quad \forall i\in\{1,...,\tau_p\}.
        \end{align*}
    \end{algsubstates}
    \item[~~\algfont{For each AP $\ell = 1, \dots, L$:}]
    \setcounter{ALC@line}{0}
    \State
    Receive the fused signals from all APs with which AP $\ell$ cooperates 
    \State\label{alg:proposed:2}
    {\algfont{For each UE $k\in\mathcal{U}_\ell$:}}
    \begin{algsubstates}
        \item\label{alg:proposed:2:e} Form the observation signals by the received fused signals from the APs $q\in\mathcal{A}_k\setminus\ell$:
        \begin{align*}
            \tilde{\bm{y}}_{\ell,k}^{p}[b,i] &\coloneqq [\bm{y}_{\ell}^{p^T}[b,i], \bm{z}_{-\ell\to\ell}^T[b,i]]^T \in \C^{P_{\ell,k}}, \forall i\in\{1,...,\tau_p\},
        \end{align*}
        where $\bm{z}_{-\ell\to\ell} \coloneqq [\bm{z}_{1\to\ell}^T, \dots, \bm{z}_{\ell-1\to\ell}^T, \bm{z}_{\ell+1\to\ell}^T, \dots, \bm{z}_{L_k\to\ell}^T]^T$
        \item\label{alg:proposed:2:f1} Derive the despread signal, $\tilde{\bm{y}}_{\ell,k}^{p,{\rm desp}}[b]$, by 
        \begin{equation*}
            \tilde{\bm{y}}_{\ell,k}^{p,{\rm desp}}[b] \coloneqq \sum_{i=1}^{\tau_p} s_k^{p}[b,i] \tilde{\bm{y}}_{\ell,k}^{p}[b,i]
        \end{equation*}
        \item\label{alg:proposed:2:f} Update the channel correlation matrices: 
        \begin{align*}
            \begin{split}
                \tilde{\bm{R}}_{\ell,k}^{p,{\rm desp}}[b] &\coloneqq \beta\tilde{\bm{R}}_{\ell,k}^{p,{\rm desp}}[b-1] + (1-\beta) \textstyle \tilde{\bm{y}}_{\ell,k}^{p, {\rm desp}}[b] \tilde{\bm{y}}_{\ell,k}^{p, {\rm desp}^H}[b]\\
                \tilde{\bm{R}}_{\ell,k}^{p}[b] &\coloneqq \beta\tilde{\bm{R}}_{\ell,k}^{p}[b-1] + (1-\beta) \textstyle \sum_{i=1}^{\tau_p} \tilde{\bm{y}}_{\ell,k}^{p}[b,i] \tilde{\bm{y}}_{\ell,k}^{p^H}[b,i] \\
                \tilde{\bm{R}}_{\ell,k}[b] &\coloneqq \frac{1}{\tau_p^2 - \tau_p} \left(\tilde{\bm{R}}_{\ell,k}^{p,{\rm desp}}[b] - \tilde{\bm{R}}_{\ell,k}^{p}[b]\right) \in \C^{P_{\ell,k} \times P_{\ell,k}}
            \end{split}
        \end{align*}
        \item\label{alg:proposed:2:g} Update the MMSE channel estimator:
        \begin{align*}
            \begin{split}
                \tilde{\bm{W}}_{\ell,k}[b] \coloneqq \tau_p \left(\tilde{\bm{R}}_{\ell,k}^{p,{\rm desp}}[b]\right)^{-1} \tilde{\bm{R}}_{\ell,k}[b] \tilde{\bm{E}}_k \in \C^{P_{\ell,k} \times N},
            \end{split}
        \end{align*}
        where $\tilde{\bm{E}}_k \coloneqq [\I_{N\times N}, \bm 0]^T \in \R^{ P_{\ell,k} \times N}$.
        \item\label{alg:proposed:2:h} Estimate the local channel $\bm{h}_{\ell,k}$:
        \begin{equation*}
            \hat{\bm{h}}_{\ell,k}[b] = \tilde{\bm{W}}_{\ell,k}^{H}[b] \tilde{\bm{y}}_{\ell,k}^{p, {\rm desp}}[b]
        \end{equation*}
    \end{algsubstates}        
\end{algorithmic}

%% file: TeX/Text/simulations.tex
In this section, the proposed cooperative CE approach is compared against the local CE approach, where APs perform CE individually, and against the centralized CE approach, where APs in each cluster send their $N$-dimensional signals to a CPU, where the CE is performed in a centralized manner. The comparison is also made against the centralized variant of the CE approach proposed in \cite{van2024distributed}, which takes into account the structure of the channel correlation matrix $\bm{R}^k$ in \eqref{eq:channel_cov} as a priori knowledge to enhance the channel correlation matrix estimation performance, and consequently, the overall CE performance. This approach, here referred to as PK-centralized, is computationally complex due to inner loop iterations. For these iterations, we set the maximum number of iteration to 5 in the simulations. Through these comparisons, we investigate the effectiveness of our proposed cooperative CE approach in terms of CE accuracy and convergence rate.

We consider a scenario where $L = 130$ APs and $K$ UEs are randomly distributed in an area of $2\times 2$ ${\rm km}^2$. Various numbers of UEs $K\in\{100, 200,300\}$ and various numbers of antennas per AP $N \in \{2,5,10\}$ are considered, with the default values $K=100$ and $N=5$. The density of APs is roughly $32$ APs$/{\rm km}^2$, while the density of UEs is $\{25,50,75\}$ UEs$/{\rm km}^2$. The formation of AP clusters $\mathcal{A}_k$ and UE subsets $\mathcal{U}_\ell$ is based on the definitions outlined in \cref{sec:AP_assignment}, with a maximum size of $|\mathcal{A}_k|=L_k=4$; \ie, each UE is served by at most 4 closest APs.

In the numerical experiments, the LoS component in the channel model \eqref{eq:channel} is defined as \cite{mukherjee_performance_2022,sun_bandwidth-efficient_2021,amadid_channel_2022}
\begin{align*}
    \mathcirc{\bm{h}}^k \coloneqq [\mathcirc{\bm{h}}_{1,k}^T&,\dots,\mathcirc{\bm{h}}_{L_k,k}^T]^T, \quad \forall k\in[K]  \numberthis \label{eq:LOS}\\
    ~~\text{with}~~ \mathcirc{\bm{h}}_{\ell,k} &= \sqrt{\beta_{\ell,k}}\sqrt{\frac{K_{\ell,k}}{K_{\ell,k}+1}} e^{j\varphi_{\ell,k}}\bm{a}_{\ell,k} \quad \forall \ell\in\mathcal{A}_k,
\end{align*}
where
\begin{align*}  
    \bm{a}_{\ell,k} &\coloneqq  [1, e^{j2\pi d_a \sin(\theta_{\ell,k})},\dots,e^{j(N-1)2\pi d_a \sin(\theta_{\ell,k})}]^T\\
    \beta_{\ell,k} &\coloneqq \rm{db2mag}(\alpha + 10\rho \log_{10} (d_{\ell,k}) + \xi), ~~ \xi \sim \mathcal{N}(0; \sigma^2_S)\\
    K_{\ell,k} &\coloneqq \rm{db2mag}(10^{1.3 - 0.003 d_{\ell,k}}).
\end{align*}
In the formulation above, $\beta_{\ell,k}$ represents the large-scale fading coefficient between UE $k$ and AP $\ell$, which models the path loss and shadowing. This factor is determined by constants $\alpha = -34$, $\rho = -3.8$, and the shadowing standard deviation $\sigma_S = 8$, and it depends on the distance $d_{\ell,k}$ (in meters) between the UE and AP. The function $\rm{db2mag}(\cdot)$ converts values in dB to linear scale. Moreover, the factor $K_{\ell,k}$ represents the ratio of signal power in the LoS component to signal power in the NLoS component. It is important to note that values of $K_{\ell,k} > 1$ indicate a stronger LoS component relative to the NLoS component, which is the case in our scenario as it is also the case due to the typical AP-UE distances considered in CFmMIMO communication networks \cite{nayebi2017precoding,buzzi2017cell,bjornson_scalable_2020}. The steering vector between UE $k$ and AP $\ell$ is denoted by $\bm{a}_{\ell,k}$, which depends on the AoA $\theta_{\ell,k}$ and on the antenna spacing in the array $d_a$ measured in number of wavelengths, and $\varphi_{\ell,k} \sim \mathcal{U}[0, 2\pi]$ indicates the signal phase at the reference antenna, \eg, the first antenna.

Furthermore, the NLoS component is defined as \cite{mukherjee_performance_2022,sun_bandwidth-efficient_2021,amadid_channel_2022}
\begin{align*}
    \dot{\bm{h}}^k[b] &\coloneqq [\dot{\bm{h}}_{1,k}^T[b],\dots,\dot{\bm{h}}_{L_k,k}^T[b]]^T, \quad \forall k\in[K] \numberthis \label{eq:NLOS}\\
    &\text{with}~~  \dot{\bm{h}}_{\ell,k}[b] = \sqrt{\beta_{\ell,k}}\sqrt{\frac{1}{K_{\ell,k}+1}} \bm{g}_{\ell,k}[b]  \quad \forall \ell\in\mathcal{A}_k,
\end{align*}
where $\bm{g}_{\ell,k}[b]$ is modeled as a correlated Rayleigh fading distribution $\bm{g}_{\ell,k}[b] \sim \mathcal{N}_{\mathcal{C}}\left(\bm 0, \E[\theta \sim \mathcal{D}^\theta]{\bm{a}_{\ell,k}(\theta)\bm{a}_{\ell,k}^H(\theta)}\right)$, where $\mathcal{D}^{\theta} = \mathcal{N}(\theta_{\ell,k}, \sigma^2_{\theta})$ with $\sigma_\theta=10^\circ$, representing small-scale fading \cite[Sec. 2.6]{bjornson_massive_2017}.
Moreover, the average SNR is set to $10$ dB unless specified otherwise. The number of pilot sequences is $\tau_p = 10$, and $\tau_d=100$ time instants are allocated for data transmission during each coherence block. A batch size of $50$ is considered for \cref{alg:CE_cooperative}, meaning that the filters $\mathcirc{\bm W}_{\ell\to j}$ and $\tilde{\bm{W}}_{\ell,k}$ are updated every $50$ channel coherence blocks, equivalent to one iteration in the depicted convergence plots. The default iteration number for all the CE approaches is set to $T=100$. In the presented cumulative density function (CDF) plots, CE is conducted with 500 different random seeds.
Finally, the CE accuracy metric is considered as CE MMSE Loss $\coloneqq \E{\nicefrac{\|\bm{h}_{\ell,k} - \hat{\bm{h}}_{\ell,k}\|^2}{\trace\{\bm{R}_{\ell,k}\}}}$.

\begin{figure}[th]
    \centering
    \includegraphics[width=.8\columnwidth]{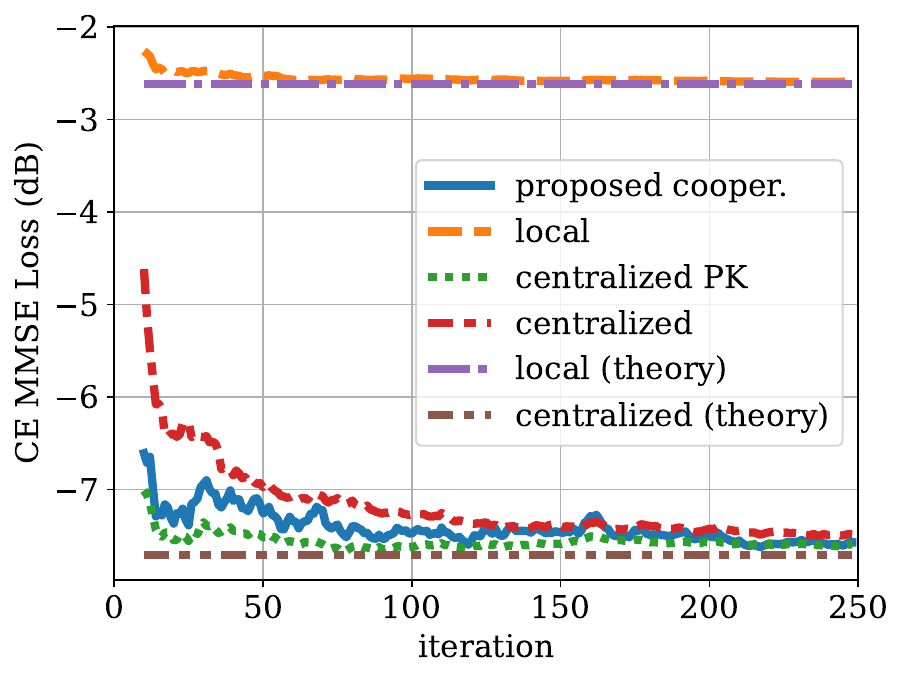}
    \caption{Convergence comparison between various CE approaches.}
    \label{fig:convergence}
\end{figure}

In the first experiment, various CE approaches are compared in terms of convergence rate for a single random seed. This comparison is depicted in \cref{fig:convergence}, where the CE MMSE loss for an AP-UE pair at the center of the considered area is plotted against the algorithm iteration number. According to the figure, the proposed cooperative CE approach converges rapidly to the theoretical value of the centralized CE approach. The convergence rate is comparable to that of the PK-centralized CE approach, which has higher computational complexity due to its inner loops. Notably, the convergence rate of the proposed approach is faster than that of the centralized CE approach. This is primarily attributed to the smaller correlation matrices $\tilde{\bm{R}}_{\ell,k}^{p,{\rm desp}} \in \C^{P_{\ell,k} \times P_{\ell,k}}$ and $\tilde{\bm{R}}_{\ell,k}^{p} \in \C^{P_{\ell,k} \times P_{\ell,k}}$ in the proposed approach compared to the centralized case with larger $\bm{R}^{p,\rm desp}_{k} \in \C^{NL_k\times NL_k}$ and $\bm{R}^{p}_k \in \C^{NL_k\times NL_k}$, which result in higher correlation estimation errors. \cref{fig:convergence} also shows the performance of the local CE approach, as well as its theoretical value, with a significant 5dB performance gap compared to the centralized CE approach in the considered case.

\begin{figure*}[th]
    \centering
    \begin{subfigure}[b]{.33\textwidth}
        \centering
        \includegraphics[width=\textwidth]{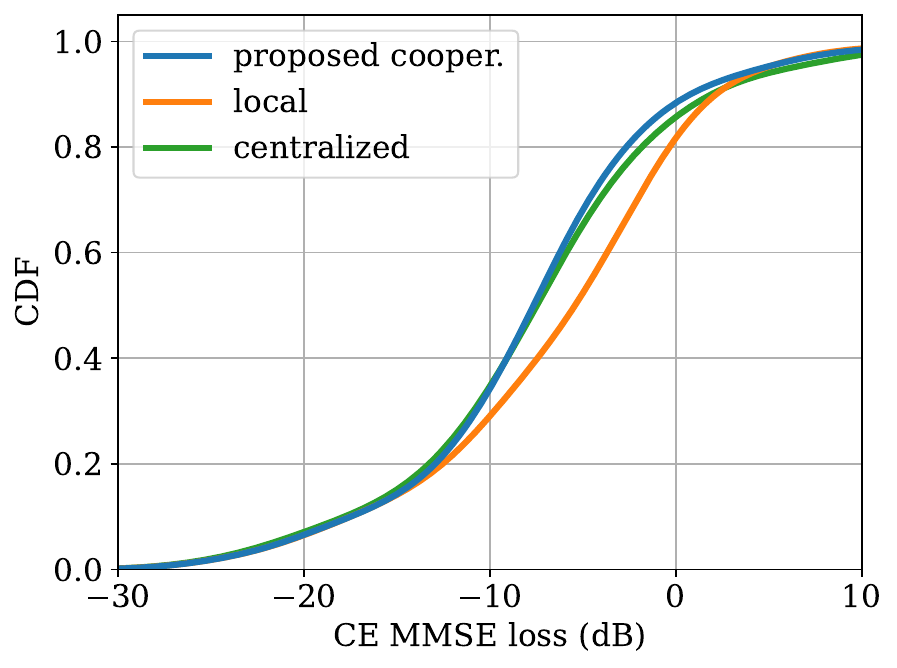}
        \caption{$T = 50$}
    \end{subfigure}%
    \begin{subfigure}[b]{.33\textwidth}
        \centering
        \includegraphics[width=\textwidth]{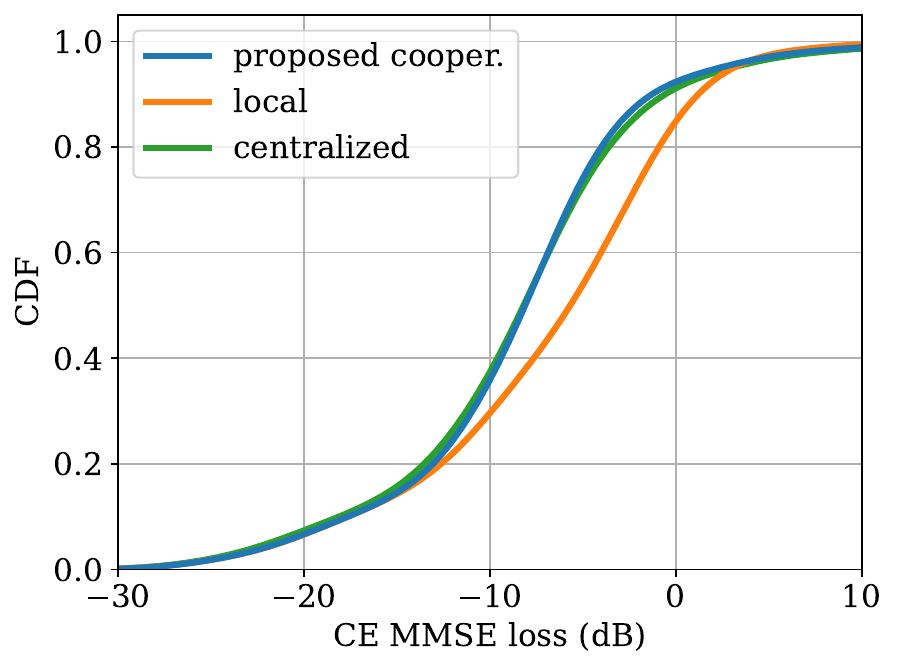}
        \caption{$T = 100$}
        \label{fig:CE_performance:T100:N5}
    \end{subfigure}%
    \begin{subfigure}[b]{.33\textwidth}
        \centering
        \includegraphics[width=\textwidth]{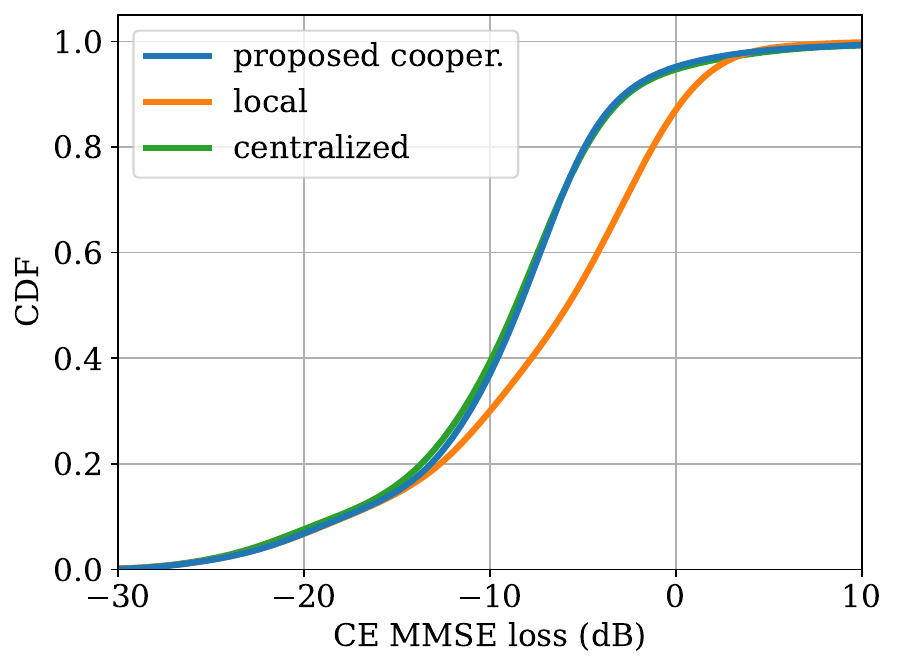}
        \caption{$T = 200$}
    \end{subfigure}
    \begin{subfigure}[b]{.33\textwidth}
        \centering
        \includegraphics[width=\textwidth]{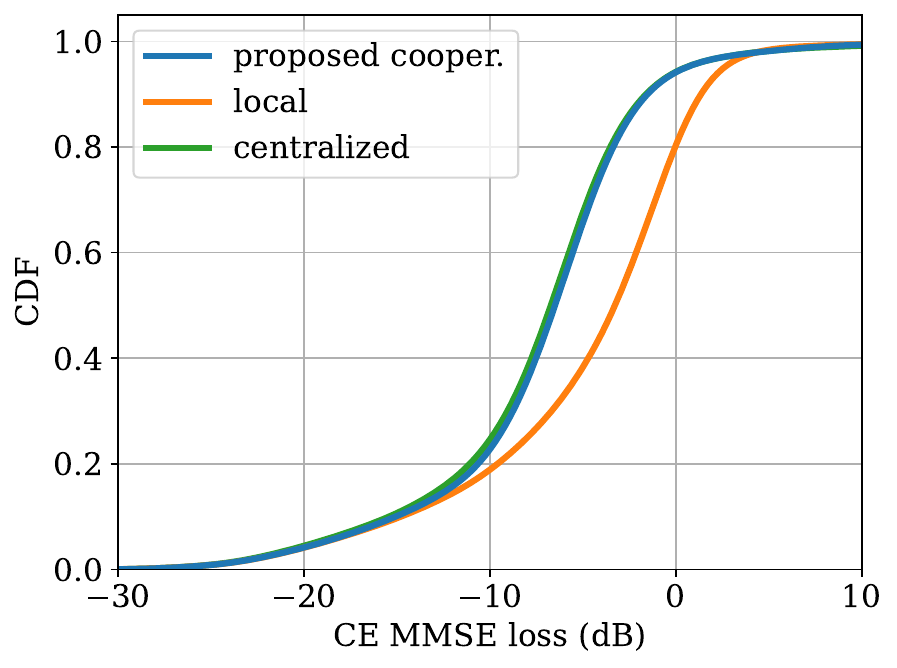}
        \caption{$N = 2$}
        \label{fig:CE_performance:N:1}
    \end{subfigure}%
    \begin{subfigure}[b]{.33\textwidth}
        \centering
        \includegraphics[width=\textwidth]{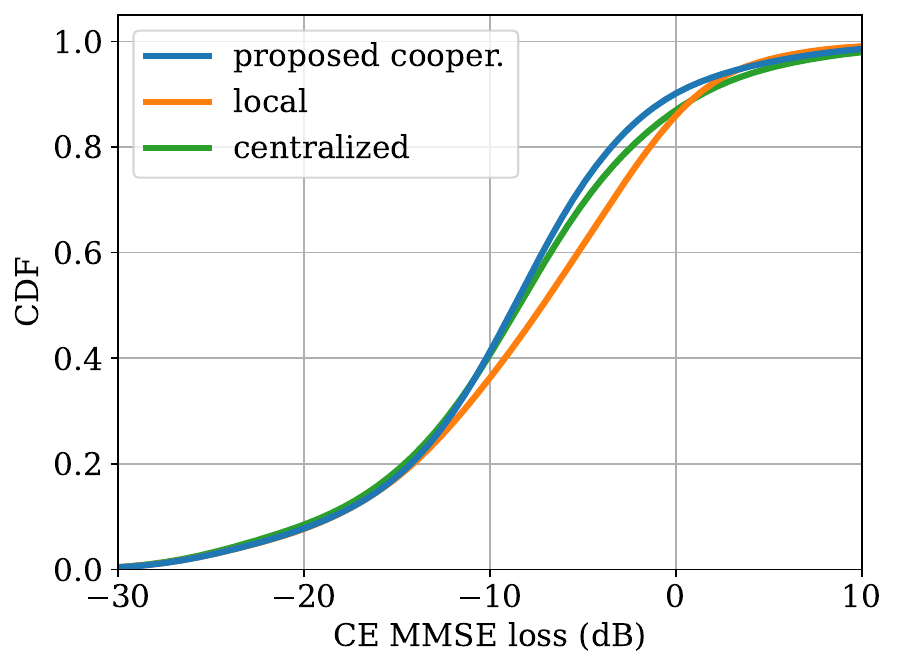}
        \caption{$N = 10$}
        \label{fig:CE_performance:N:2}
    \end{subfigure}
    \begin{subfigure}[b]{.33\textwidth}
        \centering
        \includegraphics[width=\textwidth]{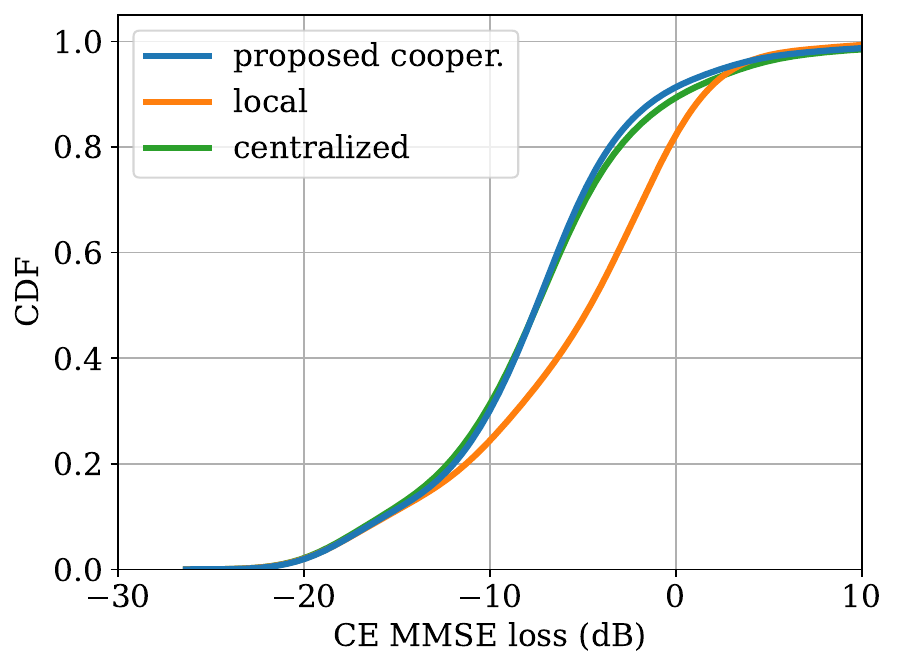}
        \caption{${\rm SNR} = 0$}
        \label{fig:CE_performance:SNR}
    \end{subfigure}%
    \caption{CE performance comparison between various CE approaches. The first row reports CE performance for different iteration numbers $T$, with the number of antennas per AP set to $N=5$ and the average SNR equal to $10$ dB. Plots (d) and (e) investigate the effect of varying the number of antennas $N$ while keeping $T=100$. Plot (f) compares CE performance at ${\rm SNR} = 0$ dB with $T=100$ and $N=5$. The number of UEs in all the plots is set to $K=100$.}
    \label{fig:CE_performance}
\end{figure*}
\begin{figure*}[th]
    \centering
    \begin{subfigure}[b]{.33\textwidth}
        \centering
        \includegraphics[width=\textwidth]{Images/CE_performance_iter100.pdf}
        \caption{$K = 100$}
    \end{subfigure}%
    \begin{subfigure}[b]{.33\textwidth}
        \centering
        \includegraphics[width=\textwidth]{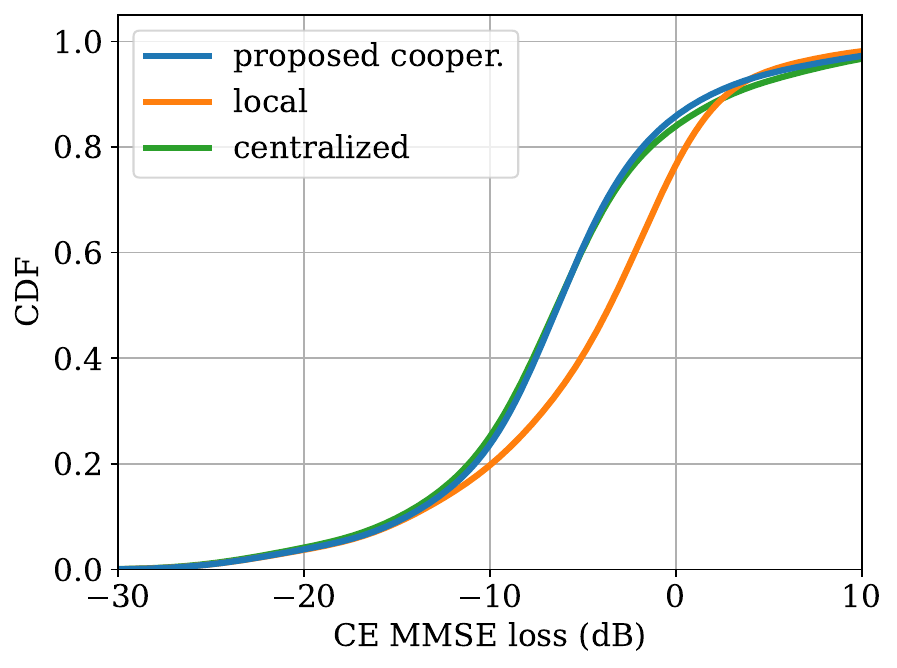}
        \caption{$K = 200$}
    \end{subfigure}%
    \begin{subfigure}[b]{.33\textwidth}
        \centering
        \includegraphics[width=\textwidth]{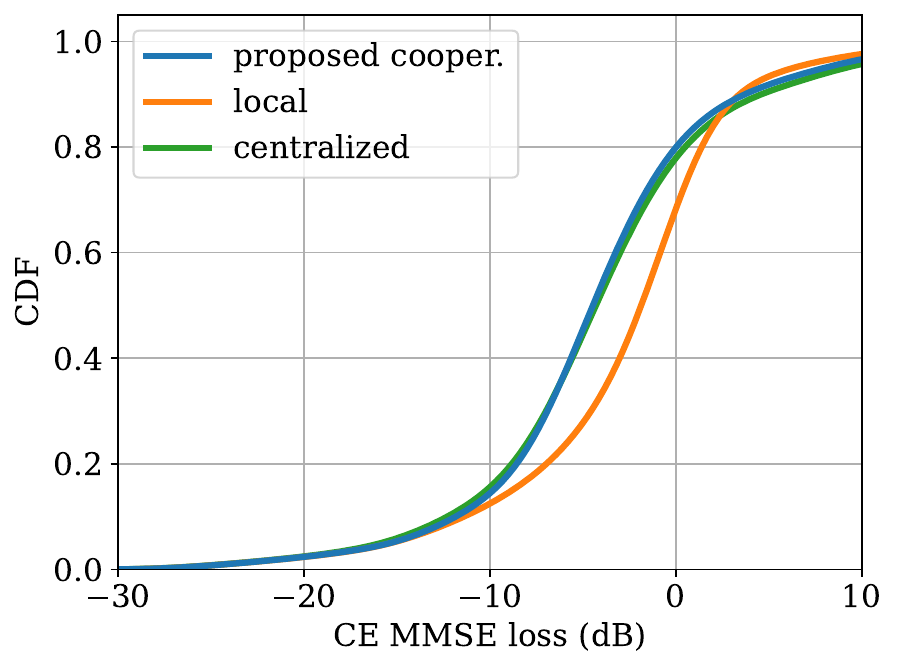}
        \caption{$K = 300$}
    \end{subfigure}
    \caption{CE performance comparison between various CE approaches with different numbers of UEs $K$ in a $2\times 2~ {\rm km^2}$ area with $T=100$ and $N=5$.}
    \label{fig:CE_performance_K}
\end{figure*}

The CDF of the CE MMSE loss is shown in \cref{fig:CE_performance} for various independent communication scenarios with different iteration numbers $T$. In the first plot with $T=50$, it is clear that the proposed approach even surpasses the centralized CE approach, with a considerable gap compared to the local CE approach. As the centralized CE approach has a slower convergence rate, increasing the number of iterations $T$ allows this approach to converge to lower CE MMSE loss values. Moreover, the performance of different CE approaches is similar for strong or weak channels, where CE MMSE loss achieves lower or higher values, respectively. The two experiments in \cref{fig:convergence,fig:CE_performance} demonstrate the superiority of the proposed CE approach in fast-changing dynamic environments, where UEs relocate rapidly, requiring fast convergence to new correlation matrices.

The next experiment is depicted in the second row of \cref{fig:CE_performance}, where the CDF of the CE MMSE loss for various independent communication scenarios is shown with different numbers of antennas per AP, denoted as $N$. In these plots, it is evident that the proposed approach achieves the same performance as the centralized CE approach and even surpasses it with a larger number of antennas per AP. This superior performance is attributed to lower estimation errors for smaller correlation matrices in the proposed approach compared to the centralized CE approach.
In this experiment, the benefit of the proposed approach is evident both in terms of CE accuracy compared to the local CE approach and in terms of communication bandwidth compared to the centralized CE approach, particularly with a larger number of antennas per AP. Furthermore, in plot \cref{fig:CE_performance:SNR}, the performance of various CE approaches is reported when the average SNR is $0$ dB. This result is comparable to \cref{fig:CE_performance:T100:N5}, which shows performance with the same number of iterations $T=100$ and antennas per AP $N=5$, but with a higher average SNR of $10$ dB.

In the last experiment, various CE approaches are compared with different numbers of UEs $K\in\{100,200,300\}$, randomly distributed in the considered area. The results are reported in \cref{fig:CE_performance_K}, where it can be concluded that all CE approaches exhibit higher performance with fewer UEs due to reduced interference. The same relative performance is observed for the considered CE approaches, with the proposed approach converging to the centralized CE approach while maintaining a considerable gap from the local CE approach.

%% file: TeX/Text/conclusion.tex
In this paper, we have proposed a cooperative CE approach for user-centric CFmMIMO communication networks. In this approach, UEs select a cluster of APs with the best channel conditions (\eg, the closest APs) to access service. Within each cluster, APs cooperate in estimating their channels to the corresponding UE. This cooperation is motivated by the higher likelihood of LoS components in CFmMIMO communication networks. The proposed cooperative CE approach is optimal, as its performance is equivalent to that of the centralized CE approach, notably by sharing fused signals rather than $N$-dimensional raw signals between APs. In the proposed cooperative CE approach, pilots are assigned randomly in each channel coherence block, eliminating the need for pilot assignment and pilot decontamination techniques. In the numerical experiments, the same CE accuracy as that of the centralized CE approach has been observed. Superior performance has been observed in terms of CE accuracy compared to the local CE approach, and in terms of convergence rate compared to the centralized CE approach.